\documentclass[journal ]{new-aiaa}
\usepackage[utf8]{inputenc}
\usepackage{textcomp}

\usepackage{graphicx}
\usepackage{amsmath}
\usepackage[version=4]{mhchem}
\usepackage{siunitx}
\usepackage{subfigmat} 
\usepackage{longtable,tabularx}
\usepackage{relsize}

\newcommand\figref[1]{Fig. \ref{fig:#1}} 
\newcommand\tabref[1]{Table \ref{tab:#1}} 
\newcommand\eref[1]{Eq. (\ref{eq:#1})} 

\newcommand\p[1]{\partial{#1}}

\title{Optimal Clipping of Structural Subgrid Stress Closures for Large Eddy Simulation}

\author{Aviral Prakash \footnote{Corresponding author. Email address: aviral.prakash@colorado.edu}, Kenneth E. Jansen, John A. Evans}
\affil{University of Colorado, Boulder, CO 80309}

\begin{document}

\maketitle

\begin{abstract}

Structural subgrid stress models for large eddy simulation often allow for backscatter of energy from unresolved to resolved turbulent scales, but excessive model backscatter can eventually result in numerical instability. A commonly employed strategy to overcome this issue is to set predicted subgrid stresses to zero in regions of model backscatter. This clipping procedure improves the stability of structural  models, however, at the cost of reduced correlation between the predicted subgrid stresses and the exact subgrid stresses.  In this article, we propose an alternative strategy that removes model backscatter from model predictions through the solution of a constrained minimization problem. This procedure, which we refer to as optimal clipping, results in a parameter-free mixed model, and it yields predicted subgrid stresses in higher correlation with the exact subgrid stresses as compared with those attained with the traditional clipping procedure. We perform a series of \textit{a priori} and \textit{a posteriori} tests to investigate the impact of applying the traditional and optimal clipping procedures to Clark's gradient subgrid stress model, and we observe that optimal clipping leads to a significant improvement in model predictions as compared to the traditional clipping procedure.

\end{abstract}

\newpage

\section*{Nomenclature}

{\renewcommand\arraystretch{1.0}
\noindent\begin{longtable*}{@{}l @{\quad=\quad} l@{}}
$u_{\tau}$  & friction velocity  \\
C.C. & correlation coefficient \\
R.E.F. & relative error in mean energy flux \\
$\Delta_i$ & filter width along $x_i$ direction\\
$\Pi$ & local exact SGS dissipation\\
$\Pi_M$ & local model dissipation\\
$\tau_{ij}$ & $ij^{th}$ component of exact SGS tensor\\
$\tau_{ij}^M$ & $ij^{th}$ component of modeled SGS tensor\\
$\eta$ & Kolmogorov length scale \\
$\delta$ & channel half width \\

\end{longtable*}}

\section{Introduction}

Simulations of turbulent flows are important for product design and development cycles in the aerospace industry. Turbulence statistics are often needed for understanding flow around a certain geometry and facilitating certification by analysis. Direct numerical simulations (DNS) resolve all turbulent scales and give accurate estimates of turbulent statistics. However, as the range of length and time scales associated with turbulent eddies increases with Reynolds number, see \citep{Chapman1979,choi2012} for estimates, DNS becomes infeasible at higher Reynolds numbers. For such flows, large eddy simulation (LES) is an attractive alternative. LES reduces the range of resolved turbulent scales by solving the filtered Navier-Stokes equations. However, filtering of the Navier-Stokes equations leads to an unclosed term, known as the subgrid-scale stress (SGS) tensor, that accounts for the interaction between resolved and unresolved turbulent scales. The SGS tensor must be modeled to arrive at a closed system of equations. Functional SGS models \citep{Smagorinsky1963, Germano1991} aim to accurately predict model dissipation, that is the transfer of energy from the resolved scales to unresolved scales. On the other hand, structural models \citep{Clark1979, Bardina1980, Stolz2001} target accurate prediction of the SGS tensor itself.  

Most of the commonly used structural models, such as Clark's gradient model \cite{Clark1979}, Bardina's scale-similarity model \cite{Bardina1980}, and approximate deconvolution models (ADM) \cite{Stolz2001}, yield SGS tensor estimates that have a high correlation with the exact SGS tensor. Unfortunately, these models often underpredict local SGS dissipation \citep{Borue1998, Liu1994}. Even for simple flows, such as homogeneous and isotropic turbulence (HIT) at high Reynolds numbers, these models lead to energy pileup at higher resolved wavenumbers. Moreover, structural models often exhibit backscatter, energy transfer from unresolved to resolved scales, and excessive model backscatter can result in numerical instabilities. This behavior has severe complications for mixing layer flows, and numerical instabilities are frequently observed while using these models \cite{Vreman1997}. A common strategy to improve the dissipative performance of structural models is to clip the modeled SGS tensor whenever backscatter is present \cite{Liu1994}. Another method to improve the dissipative performance of SGS models is to use mixed models that are defined as a linear combination of structural models and functional models \citep{Clark1979, Vreman1997}. A common mixed model is a combination of Clark's gradient model \cite{Clark1979} and the classical Smagorinsky model \cite{Smagorinsky1963}. This model avoids numerical instability, however, the model is overdissipative and performs poorly for laminar-turbulent transition \cite{Vreman1996}. Combining Clark's gradient model with the dynamic Smagorinsky model \cite{Germano1991} results in a so-called dynamic mixed model \citep{Zang1993, Vreman1996} and gives better dissipative performance. Similar observations also hold for other structural models such as Bardina's scale-similarity model \cite{Vreman1997}. Another method to improve the dissipative performance of Clark's gradient model is dynamic regularization \cite{Vollant2016}. In this method, the gradient model form is decomposed into several components, and the terms that contribute to model backscatter are eliminated. All of the above strategies to improve the prediction of model dissipation share one common drawback: they reduce the correlation between the modeled and the true SGS tensor or the so-called structural accuracy of the underlying structural model. For anisotropic flows at moderate grid resolutions, accurate estimation of the SGS tensor is essential for accurate predictions of turbulent statistics \cite{Baggett1997} and therefore, methods that ensure a high structural accuracy are needed. 

In this article, we propose a novel strategy to improve the dissipative performance of a structural model that has minimal impact on the model's structural accuracy. The strategy involves the solution to a constrained optimization problem that ensures that the difference between the modified and original SGS model is minimized while satisfying the constraint of no local model backscatter. We refer to this strategy as optimal clipping. Application of optimal clipping to a structural SGS model yields a parameter-free mixed model that has a lower model evaluation cost compared to a dynamic mixed model. In Section \ref{sec:Methodology}, we first discuss the standard clipping procedure before presenting our new optimal clipping procedure. In Section \ref{sec:Results}, we perform \textit{a priori} and \textit{a posteriori} tests to assess the performance of optimally clipping as applied to Clark's gradient model and compare the results to those from other methodologies. In Section \ref{sec:conclusions}, we summarize the results presented in this article and propose directions for future research.

\section{Methodology}
\label{sec:Methodology}

In LES, one solves the filtered Navier-Stokes equations for the filtered flow field:

\begin{equation}
    \frac{\partial \bar{u}_i}{\partial x_i} = 0,
\end{equation}

\begin{equation}
\frac{\partial \Bar{u}_i}{\p t} + \frac{\partial}{\p x_j} (\Bar{u}_i \Bar{u}_j)    = - \frac{1}{\rho} \frac{\partial \Bar{p}}{\p x_i} + \frac{\partial }{\p x_j} (2 \nu \Bar{S}_{ij}) - \frac{\p \tau_{ij}}{\p x_j} + \bar{f}_i, \qquad \tau_{ij} = \overline{u_i u_j} - \Bar{u}_i \Bar{u}_j,
\end{equation}

\noindent where $\bar{u}_i$ is the filtered velocity component along the $i^{th}$ direction, $\bar{p}$ is the filtered pressure, $\bar{S}_{ij} = \frac{1}{2} \Big( \frac{\partial \Bar{u}_i}{\partial x_j} + \frac{\partial \bar{u}_j}{\partial x_i} \Big)$ is the filtered strain-rate tensor, and $\tau_{ij}$ is the so-called subgrid scale stress (SGS) tensor. The SGS tensor accounts for the interaction between the filtered (resolved) flow field and the sub-filter (unresolved) flow field. This tensor depends on the unresolved velocity field which is unknown during the simulation. Consequently, the SGS tensor must be approximated with a model in practice.

The most commonly used SGS models are based on the gradient-viscosity hypothesis. One of the earliest models is the Smagorinsky model \cite{Smagorinsky1963} which models the deviatoric part of the stress tensor as:

\begin{equation}
    \tau_{ij}^{M,d} = \tau_{ij}^M - \frac{1}{3} \delta_{ij} \tau_{kk}^M = -2 \nu_t \bar{S}_{ij}, \qquad \nu_t = (C_s \Delta)^2 \vert \bar{\pmb{S}} \vert,
\end{equation}

\noindent where $\tau_{ij}^{M}$ is the modeled SGS tensor, $\tau_{ij}^{M,d}$ is the deviatoric part of the modeled SGS tensor, $C_s$ is the Smagorinsky constant, and $\Delta$ is the filter width. The filter width is commonly taken to be $\Delta = (\Delta_1 \Delta_2 \Delta_3)^{1/3}$, where $\Delta_i$ is the grid spacing in the $i^{th}$ direction \cite{Deardorff1970}. The optimal value of the Smagorinsky constant depends on the flow under consideration \citep{Lilly1992, Bardina1980, Canuto1997}. The dynamic procedure introduced by Germano \citep{Germano1991, Germano1992, Lilly1992} uses a coarse grid test-filter to approximate the value of the optimal Smagorinsky constant, thereby reducing user dependence for different flows. The anisotropic minimum dissipation (AMD) model \cite{Rozema2015} instead employs the minimum eddy-viscosity needed to dissipate the unresolved scales. Despite their differences, the SGS model form for all these models is based on the assumption of local and instantaneous alignment between the deviatoric part of the SGS tensor and the resolved strain-rate tensor. However, statistical \textit{a priori} studies have demonstrated this stress-strain alignment assumption generally does not hold \citep{Clark1979, Liu1994}, and SGS models which are based on the gradient-viscosity hypothesis are characterized by a low correlation between the true and predicted SGS tensors. It was mentioned in \cite{Baggett1997} that an inaccurate stress representation does not lead to an incorrect prediction of turbulent statistics like energy spectra for HIT as long as the model dissipation is accurately predicted. However, for anisotropic flows like wall-bounded flows where the flow is driven by large turbulent eddies that dominate shearing, a finer mesh resolution is required to get accurate results \cite{Baggett1997}. For such flows, SGS models with improved structural accuracy can reduce the mesh requirement needed for accurate prediction of turbulence statistics.

There are other SGS models such as Clark's gradient model \citep{Clark1979,Borue1998}, Bardina's scale-similarity model \cite{Bardina1980}, and the approximate deconvolution model (ADM) \cite{Stolz2001} that do not rely on the assumption of stress-strain alignment. Instead, these models rely on mathematical techniques such as formal series expansion to accurately predict the SGS tensor. As these models aim to accurately predict the SGS tensor itself, they are referred to as structural SGS models. \textit{A priori} statistical studies have shown that structural SGS models generally yield SGS tensor predictions that are in high correlation with the true SGS tensor \citep{Liu1994,Borue1998}, and consequently, they have a high structural accuracy. Another advantage of these models is that the local model dissipation, $\Pi_M = - \tau^{M}_{ij} \bar{S}_{ij}$, and the local energy flux based on filtered DNS data, $\Pi = - \tau_{ij} \bar{S}_{ij}$, also have a high correlation \citep{Liu1994, Borue1998}. Unlike gradient-viscosity hypothesis based models which normally enforce a non-negative eddy viscosity, structural models are capable of predicting backscatter, transfer of energy from unresolved scales to resolved scales. However, even though the exact and model dissipation are often highly correlated for structural SGS models, the exact dissipation is often under-predicted for these models leading to energy build-up at the smallest resolved scales. Moreover, model backscatter can also negatively influence the numerical stability and convergence of simulations \citep{Liu1994, Vreman1996}.
To see this, note that the energy associated with an LES satisfies the following balance law when subject to homogeneous or periodic boundary conditions and no external forcing \cite{Peters2020}:

\begin{equation}
    \frac{d}{dt} \int_{\Omega} \frac{1}{2} \bar{u}_i^2 d\Omega + \int_{\Omega} (2\nu \bar{S}_{ij} - \tau_{ij}) \bar{S}_{ij} d\Omega = 0,
\end{equation}

\noindent where $\Omega \in \mathbb{R}^3$ is the domain of interest. Decay of filtered kinetic energy is thus ensured if and only if

\begin{equation}
    \int_{\Omega} (2\nu \bar{S}_{ij} - \tau_{ij}) \bar{S}_{ij} d\Omega \geq 0
    \label{eq:int_eq}
\end{equation}

\noindent or, equivalently,

\begin{equation}
    \int_{\Omega} \tau_{ij} \bar{S}_{ij} d\Omega \leq \int_{\Omega} 2 \nu \bar{S}_{ij} \bar{S}_{ij}.
    \label{eq:int_eq_2}
\end{equation}

\noindent The above global inequality obviously holds if the local inequality $\tau_{ij} \bar{S}_{ij} \leq 2 \nu \bar{S}_{ij} \bar{S}_{ij}$ holds. However, if $\int_{\Omega} \tau_{ij} \bar{S}_{ij} d \Omega$ is sufficiently large, the inequality does not hold. This indicates that excessive model backscatter can lead to an increase in turbulent kinetic energy in time which is detrimental to the numerical stability. 

Several strategies to overcome this issue have been considered in the literature \citep{Liu1994, Vreman1996, Zang1993, Vollant2016}. One approach is to supplement a structural SGS model with a Smagorinsky model like term, resulting in a mixed SGS model 

\begin{equation}
    \tau^{M-{mixed}}_{ij} = \tau_{ij}^M - 2 \nu_t S_{ij}, \qquad \nu_t = (C_G \Delta)^2 \vert \bar{\pmb{S}} \vert.
\end{equation}

\noindent where $\tau_{ij}^M$ is the SGS tensor associated with the original SGS model, $\tau_{ij}^{M-{mixed}}$ is the SGS tensor associated with the mixed SGS model, and $C_G$ is a model constant. This approach was first proposed in \cite{Clark1979} to improve the stability of Clark's gradient model. In mixed SGS models of the above form, the constant $C_G$ must be carefully tuned to ensure accurate predictions of turbulence statistics. If $C_G$ is chosen to be too large, the resulting model is overdissipative leading to inaccurate flow predictions during laminar-turbulent transition \cite{Vreman1996}. Alternatively, if $C_G$ is chosen to be too small, the resulting model is numerically unstable. This behavior was improved in \citep{Zang1993, Vreman1996} by using the dynamic procedure \cite{Germano1991} to determine the value of $C_G$ and the resulting models are commonly known as dynamic mixed models.

Another strategy to improve numerical stability is by avoiding model backscatter, that is, to enforce

\begin{equation}
 \tau_{ij} \bar{S}_{ij} \leq 0
\end{equation}

\noindent in a pointwise manner. An obvious way to ensure this inequality holds is by clipping \cite{Liu1994}. Clipping sets stresses locally to zero when local backscatter is predicted by the model. Mathematically, the formulation for clipping can be expressed as, 

  \begin{equation}
     \tau^{M-SC}_{ij} = \begin{cases}
     0 & \textup{ if } -\tau^{M}_{ij} \bar{S}_{ij} < 0 \\
     \tau^{M}_{ij} & \textup{ if } -\tau^{M}_{ij} \bar{S}_{ij} \geq 0 \\
     \end{cases}
 \end{equation}
 
\noindent where $\tau_{ij}^M$ is again the SGS tensor associated with the original SGS model and $\tau_{ij}^{M-SC}$ is the clipped SGS tensor. We refer to the above clipping procedure as standard clipping to distinguish it from the new clipping procedure proposed in this paper. As evident from the formulation, the increased model dissipation due to standard clipping comes at the cost of a decrease in correlation between the true and predicted SGS tensors.

In this paper, we propose a clipping formulation that has minimal effect on structural accuracy while simultaneously removing model backscatter. We refer to this as optimal clipping. To prevent backscatter while simultaneously maintaining structural performance, we seek the stress tensor which best fits the structural model in terms of Frobenius norm error within the set of models not admitting local backscatter. This yields the model form

\begin{equation}
    \tau^{M-OC}_{ij} = \underset{\Tilde{\tau} \in \mathcal{S}}{\text{arg min}} \frac{1}{2} \sum_{i,j} \Big( \Tilde{\tau}_{ij} - \tau^{M}_{ij} \Big)^2 ,
    \label{eq:minprob}
\end{equation}

\noindent where $\tau_{ij}^M$ is the SGS tensor associated with the original SGS model and $\tau_{ij}^{M-OC}$ is the optimally clipped SGS tensor and $\mathcal{S}$ is the set of models not admitting local backscatter defined as 

\begin{equation}
    \mathcal{S} := \Big\{ \Tilde{\pmb{\tau}} \in \text{Sym} (3) : \tilde{\tau}_{ij} \bar{S}_{ij} \leq 0 \Big\},
\end{equation}

 \noindent where $\text{Sym} (3)$ is the space of symmetric rank-2 tensors in $\mathbb{R}^3$. If $\pmb{\tau}^{M} \in \mathcal{S}$, then $\pmb{\tau}^{M-OC} = \pmb{\tau}^{M}$. If $\pmb{\tau}^{M} \not\in \mathcal{S}$, the minimization problem in \eref{minprob} has a unique solution, and the component wise values of $\tau_{ij}^{M-OC}$ can be obtained by solving the following linear system of equations:

\begin{equation}
    \begin{bmatrix}
    1 & 0 & 0 & 0 & 0 & 0 & \bar{S}_{11} \\
    0 & 1 & 0 & 0 & 0 & 0 & \bar{S}_{22} \\
    0 & 0 & 1 & 0 & 0 & 0 & \bar{S}_{33} \\
    0 & 0 & 0 & 2 & 0 & 0 & 2\bar{S}_{12} \\
    0 & 0 & 0 & 0 & 2 & 0 & 2\bar{S}_{13} \\
    0 & 0 & 0 & 0 & 0 & 2 & 2\bar{S}_{23} \\
    \bar{S}_{11} & \bar{S}_{22} & \bar{S}_{33} & 2 \bar{S}_{12} & 2 \bar{S}_{13} & 2 \bar{S}_{23} & 0
\end{bmatrix}\begin{bmatrix}
\tau^{M-OC}_{11} \\ \tau^{M-OC}_{22} \\ \tau^{M-OC}_{33} \\ \tau^{M-OC}_{12} \\ \tau^{M-OC}_{13} \\ \tau^{M-OC}_{23} \\ \lambda 
\end{bmatrix}=\begin{bmatrix}
\tau^{M}_{11} \\ \tau^{M}_{22} \\ \tau^{M}_{33} \\ 2 \tau^{M}_{12} \\ 2 \tau^{M}_{13} \\ 2 \tau^{M}_{23} \\  0
\end{bmatrix}
\label{eq:clip_matrix}
\end{equation}
 
\noindent where $\lambda$ is the Lagrange multiplier associated with the constraint $-\tau_{ij}^{M-OC} \bar{S}_{ij} = 0$. The special structure of the system of equations enables one to easily derive

\begin{equation}
    \lambda = \frac{ \tau^{M}_{ij} \bar{S}_{ij} }{\bar{S}_{ij}\bar{S}_{ij}}.
\end{equation}

 \noindent Using this value of $\lambda$, the SGS tensor components of the optimally clipped model, $\tau_{ij}^{M-OC}$, can be obtained via sparse, explicit equations from \eref{clip_matrix}, yielding,
 
\begin{equation}
    \tau_{ij}^{M-OC} = \tau_{ij}^{M} - \lambda \bar{S}_{ij}. 
\end{equation}

\noindent Note that the above formula holds for both $\tau^M \in S$ and $\tau^M \not\in S$ if we set:

\begin{equation}
    \lambda = -\min \Bigg\{ \frac{ -\tau^{M}_{ij} \bar{S}_{ij} }{\bar{S}_{ij}\bar{S}_{ij}},0 \Bigg \}.
\end{equation}

Interestingly, the model form resulting from optimal clipping is that of a mixed model whose eddy viscosity is equal to half of the Lagrange multiplier. This value of eddy viscosity is precisely the one required by mixed models to avoid model backscatter and recover the underlying structural SGS model when there is no model backscatter. Note the optimally clipped model smoothly transitions from the original SGS model when there is no model backscatter to a mixed SGS model when there is model backscatter. The traditional approach of deriving a mixed model starts with the assumption of a mixed model form and then determining the coefficients of the model form based on physical or mathematical constraints. In our approach, a mixed model form was not initially assumed and instead resulted from the solution of the constrained minimization problem associated with optimal clipping. This in turn provides some mathematical justification for the use of mixed SGS models. We observe that optimal clipping has some additional desirable characteristics: 

\begin{itemize}
    \item The Lagrange multiplier is determined using the local flow state, and therefore, unlike a dynamic mixed model, optimal clipping does not require additional filtering or averaging over homogeneous directions. This significantly reduces the model evaluation cost.
    
    \item Although it has been observed that including Smagorinsky-like model terms in a mixed SGS model has a small impact on structural accuracy provided the Smagorinsky constant is chosen sufficiently \citep{Vreman1996,Liu1994}, it is not guaranteed mathematically. Furthermore, averaging over homogeneous directions can reduce the structural accuracy. Alternatively, with optimal clipping, we ensure that structural accuracy is minimally affected.
    
    \item Like other mixed model formulations \cite{Silvis2017}, optimal clipping preserves the near-wall scaling of the underlying structural SGS model.
    
\end{itemize}

\noindent In this article, we will demonstrate the applicability of optimal clipping by using it in conjunction with Clark's gradient model \cite{Clark1979}. However, it must be highlighted that the optimal clipping formulation can be easily used with other models: $\tau^M_{ij}$ can be defined by Bardina's scale-similarity model \cite{Bardina1980}, the approximate deconvolution model \cite{Stolz2001} or even a data-driven SGS model \cite{Prakash2021}. 

\section{Results}

\label{sec:Results}

We consider Clark's gradient model \cite{Clark1979} for demonstrating the performance of optimal clipping. Clark's gradient model is formulated by performing a Taylor-series expansion in Fourier space, truncating higher-order terms, and transforming back to the physical space. If we use a filter kernel corresponding to an isotropic box-filter, we attain the following model form \cite{Clark1979}:

\begin{equation}
    \tau^M_{ij} = \frac{1}{12} \Delta^2 \frac{\partial \bar{u}_i}{\partial x_k} \frac{\partial \bar{u}_j}{\partial x_k},
\end{equation}

\noindent where $\Delta$ is the filter width taken as per Smagorinsky model definition \cite{Deardorff1970}. For an anisotropic box-filter, we attain \citep{Vreman1996, Trias2017}:

\begin{equation}
    \tau^M_{ij} = \frac{1}{12} G_{ik} A_{kl} G_{jm} A_{ml},
    \label{eq:gradmodel}
\end{equation}

\noindent where $G_{ij} = \p u_i/ \p x_j$ is the velocity gradient tensor and $\text{A}_{ij}$ is the anisotropy tensor that accounts for anisotropy of the filter. For an anisotropic box-filter aligned with the coordinate axis, the anisotropy tensor components can be written in matrix form as

\begin{equation}
    \begin{bmatrix} A_{ij} \end{bmatrix} = \begin{bmatrix}
    \Delta_1 & 0 & 0 \\
    0 & \Delta_2 & 0 \\
    0 & 0 & \Delta_3
    \end{bmatrix},
\end{equation}

\noindent where $\Delta_i$ is the filter width along the $i^{th}$ coordinate direction. The implementation of the dynamic mixed model with Clark's gradient model as the structural model requires an explicit definition of $\Delta$ in the Smagorinsky-like term. However, the optimal definition of $\Delta$ for anisotropic grids is not yet known \cite{Trias2017}. Therefore, we modify the implementation by applying the dynamic procedure to directly obtain the value of $(C_G \Delta)^2$.

In this section, we test the performance of the optimally clipped gradient model by comparing it against several SGS models. We consider the following implicit or explicit SGS models in this article: (i) an implicit SGS model or no-model (NM) whose details are mentioned in Section \ref{sec:Apost}, (ii) the dynamic Smagorinsky model (DS), (iii) the gradient model with no clipping (GM-NC), (iv) the gradient model with standard clipping (GM-SC), (v) the gradient model with optimal clipping (GM-OC), and (vi) the dynamic mixed model based on the gradient model (GM-DMM). For GM-NC, GM-SC, and GM-OC, we first perform \textit{a priori} tests, that is, we compare the modeled and exact SGS tensors obtained from isotropically and anisotropically filtered DNS data. These tests allow us to assess the impact of clipping on the prediction of stresses and model dissipation. Then, we perform \textit{a posteriori} tests which involve using all models under consideration in LES and investigating their impact on mean flow predictions and turbulence statistics. We perform \textit{a posteriori} tests for three different flows: (i) forced HIT at a high Reynolds number, (ii) Taylor-Green Vortex flow at $Re = 1600$, and (iii) turbulent channel flow at $Re_{\tau} = 590$.

\subsection{\textit{A priori} tests}

\textit{A priori} tests are conducted based on forced HIT DNS data at $Re_{\lambda} = 418$ extracted from the John Hopkins Turbulence Database (JHTDB) \cite{Li2008}. For both isotropically and anisotropically filtered DNS data, the box-filter kernel is used. The isotropically box-filtered data is extracted using the data-analysis tool offered by JHTDB, while anisotropically box-filtered data is determined by filtering the data obtained from the online data extraction service.

\subsubsection{Isotropically filtered DNS at $Re_{\lambda} = 418$}

Isotropically filtered DNS data based on several filter widths is obtained from JHTDB. Details of the data are given in \tabref{Apriori_descp}. The extracted filtered velocity gradient is used to evaluate the tensor predicted by the gradient SGS model, and this tensor is compared with the true SGS tensor extracted from the JHTDB database. This comparison is done based on two quantities: the mean correlation coefficient (C.C.) and the relative error in mean energy flux (R.E.F.). These two quantities are defined as:

\begin{equation}
    \text{C.C.} = \mathlarger{\sum_{i}} \mathlarger{\sum_{j}} \frac{\langle ( \tau_{ij}- \langle \tau_{ij} \rangle) ( \tau^M_{ij} - \langle \tau^M_{ij} \rangle) \rangle}{(\langle \tau_{ij}- \langle \tau_{ij} \rangle)^2  \rangle)^{1/2}(\langle \tau^M_{ij}- \langle \tau^M_{ij} \rangle)^2  \rangle)^{1/2}},
    \label{eq:CC}
\end{equation}

\begin{equation}
    \text{R.E.F.} = \frac{\langle \Pi^M \rangle - \langle \Pi \rangle }{\langle \Pi \rangle}
    \label{eq:REF}.
\end{equation}

\noindent Above, $\langle \cdot \rangle$ denotes spatial averaging performed over a time-step. C.C. is an estimate of the structural accuracy of the modeled stresses, whereas R.E.F. quantifies the dissipative performance of the model. A perfect SGS model has C.C. $= 1$ and R.E.F. $= 0$.

\begin{table}[b!]
    \centering
    \begin{tabular}{cccc}
        \hline
        \hline
         \textbf{No. of Samples} & \textbf{Spatial Locations} &\textbf{Time} & \textbf{Filter widths} \\
         \hline
         384 x 384 x 1 & Uniformly Sampled & $t$ = 1 & $\Delta \approx 6.5 \eta - 68 \eta $ \\
         & From Slice z = $\pi$ & & \\
         \hline
         \hline
    \end{tabular}
    \caption{Description of dataset for \textit{a priori} tests based on isotropically filtered DNS data}
    \label{tab:Apriori_descp}
\end{table}

\begin{figure}[t!]
    \centering
    \subfigure[\label{fig:Apriori_Iso_CC}]{\includegraphics[width=0.49\textwidth]{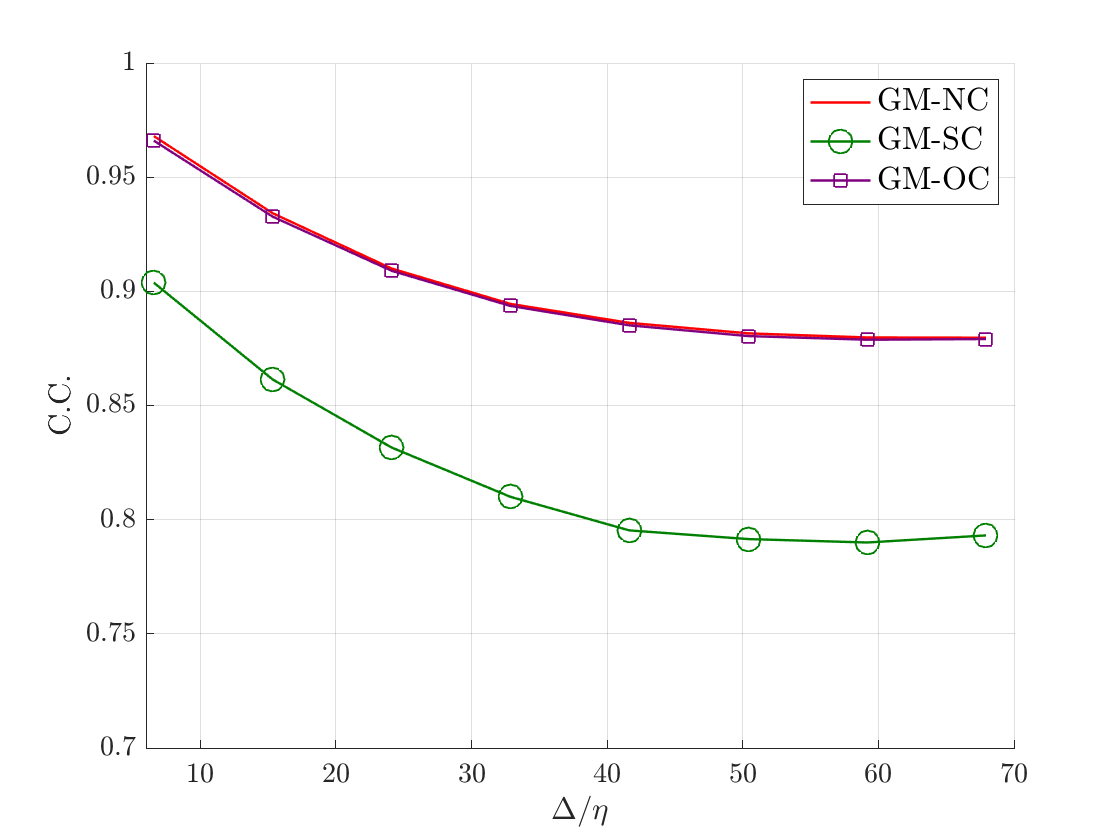}}
    \subfigure[\label{fig:Apriori_Iso_dissip}]{\includegraphics[width=0.49\textwidth]{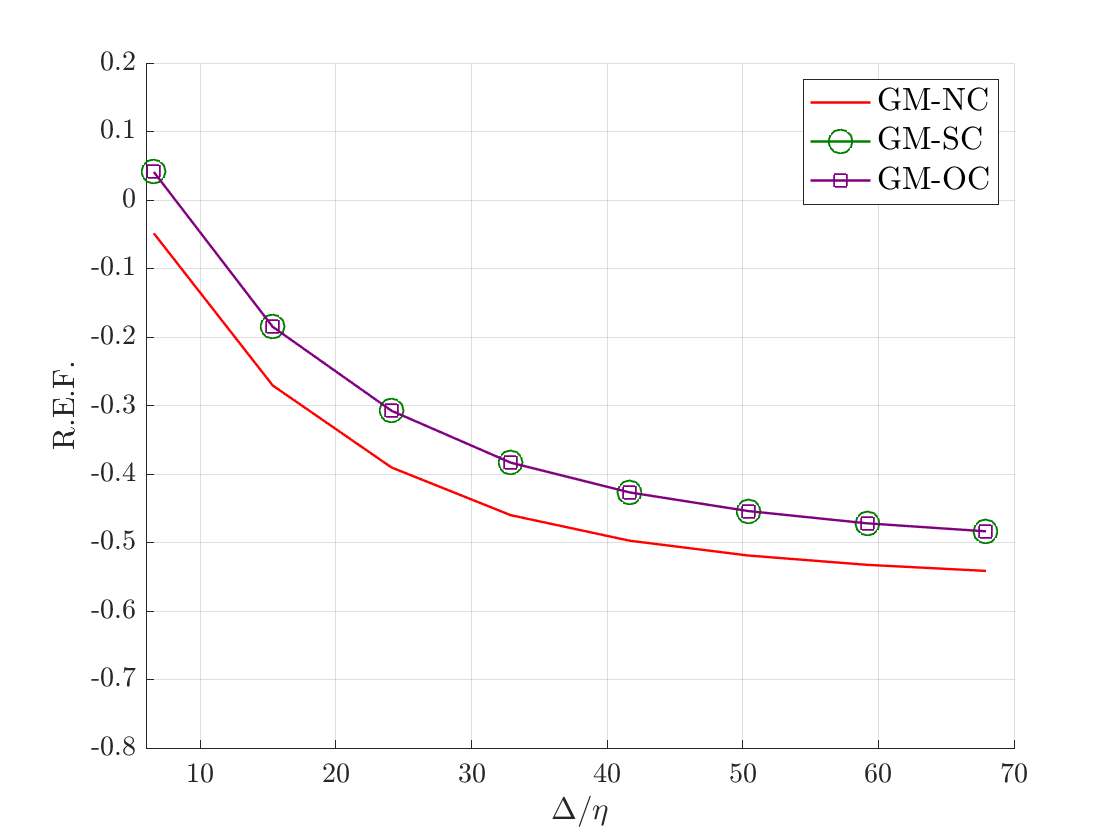}}
    \caption{\textit{A priori} results for isotropically filtered data: (a) Correlation Coefficient (C.C.) and (b) Relative Error in Mean Energy Flux (R.E.F.)}
    \label{fig:Apriori_Iso}
\end{figure}

\figref{Apriori_Iso} shows the impact of standard and optimal clipping on C.C. and R.E.F.. We observe that the SGS tensor predicted by the unclipped gradient model is highly correlated with that obtained from filtered DNS for all considered filter sizes. However, the unclipped gradient model also significantly underpredicts the SGS dissipation, that is, R.E.F. is a negative value, which manifests as reduced stability in \textit{a posteriori} simulations. The application of standard clipping to the gradient model improves the prediction of the SGS dissipation, however, this comes at the cost of reduced correlation between the predicted SGS tensor and the exact SGS tensor. The optimally clipped gradient model on the other hand preserves the high correlation offered by the unclipped gradient model and at the same time gives a SGS dissipation prediction identical to that of the standard clipped gradient model.

\subsubsection{Anisotropically filtered DNS at $Re_{\lambda} = 418$}

The isotropically filtered DNS data used in the previous section is well suited to assess the impact of clipping for cases where the turbulence is resolved using isotropic grids as is commonly done in the case of HIT. However, for common flows such as turbulent channel or a boundary layer flow, computational limitations require the use of anisotropic grids based on book-type elements, that is hexes with two long edges and a third shorter edge, in the boundary layer region. Similarly, pencil-type elements, that is hexes with two short edges and a third longer edge, can also be used in meshes involving complex geometries. The effect of clipping must be investigated for such cases, and therefore, in this section, we conduct \textit{a priori} tests using anisotropic filters that imitate the use of anisotropic grids. In particular, DNS data from JHTDB is anisotropically filtered for \textit{a priori} tests. Details about the data and filtering are given in \tabref{Apriori_descp_aniso}.

\begin{table}[t]
    \centering
    \begin{tabular}{ccccccc}
        \hline
        \hline
         \textbf{Case} & \textbf{No. of Samples} & \textbf{Spatial Locations} &\textbf{Time} & \multicolumn{3}{c}{\textbf{Filter widths}} \\
         & & & & $\Delta_1$ & $\Delta_2$ & $\Delta_3$ \\
         \hline
         Book & 64 x 64 x 64 & Randomly Sampled & $t$ = 1 & $15 \eta$ & $ 15 \eta$ & $15 \eta$ \\
          & & from $(0.5 \pi, 1.5 \pi)^3$& & $15 \eta$ & $77 \eta$  & $77 \eta$ \\         
          & & & & $15 \eta$ & $169 \eta$ & $169 \eta$ \\         
          & & & & $15 \eta$ & $231 \eta$ & $231 \eta$ \\         
          & & & & $15 \eta$ & $292 \eta$ & $292 \eta$ \\        
         Pencil & 64 x 64 x 64 & Randomly Sampled & $t$ = 1 & $15 \eta$ & $ 15 \eta$ & $15 \eta$ \\
          & & from $(0.5 \pi, 1.5 \pi)^3$ & & $15 \eta$ & $15 \eta$  & $77 \eta$ \\         
          & & & & $15 \eta$ & $15 \eta$ & $169 \eta$ \\         
          & & & & $15 \eta$ & $15 \eta$ & $231 \eta$ \\         
          & & & & $15 \eta$ & $15 \eta$ & $292 \eta$ \\                  

         \hline
         \hline
    \end{tabular}
    \caption{Description of dataset for \textit{a priori} tests based on anisotropically filtered DNS data}
    \label{tab:Apriori_descp_aniso}
\end{table}

\begin{figure}[ht!]
    \centering
    \subfigure[\label{fig:Apriori_aniso_CC_book}]{\includegraphics[width=0.49\textwidth]{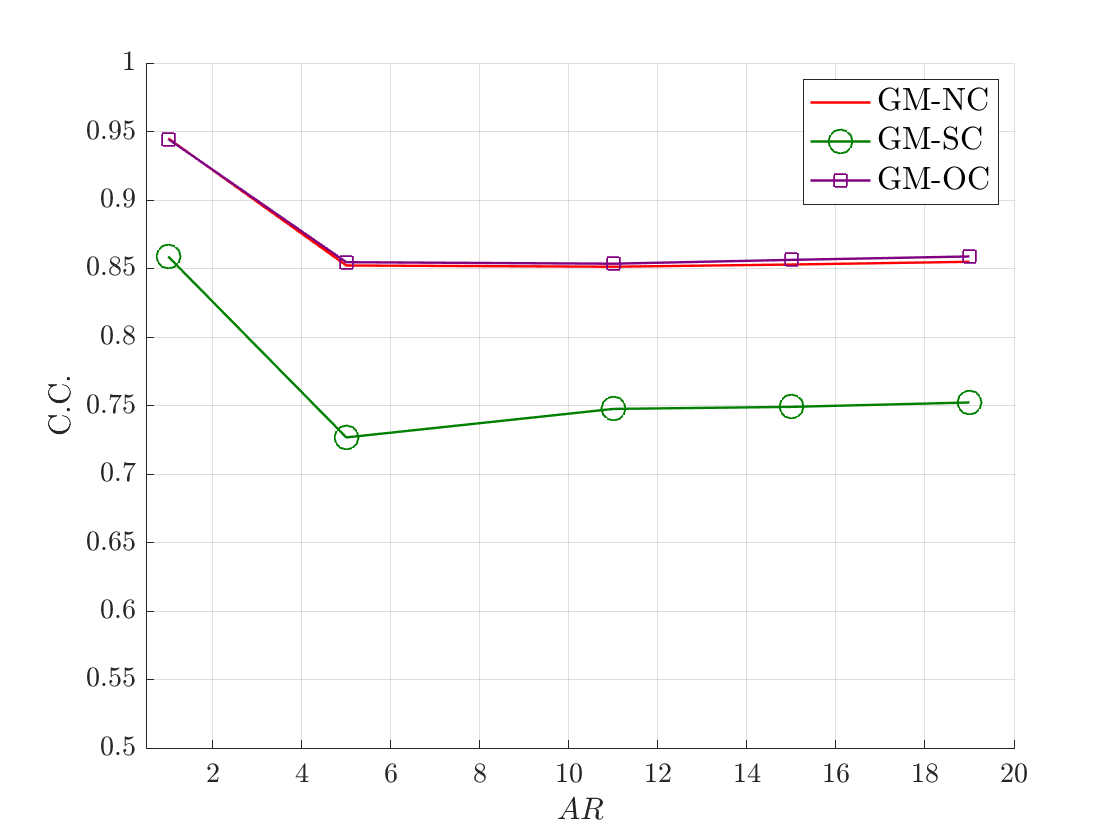}}
    \subfigure[\label{fig:Apriori_aniso_dissip_book}]{\includegraphics[width=0.49\textwidth]{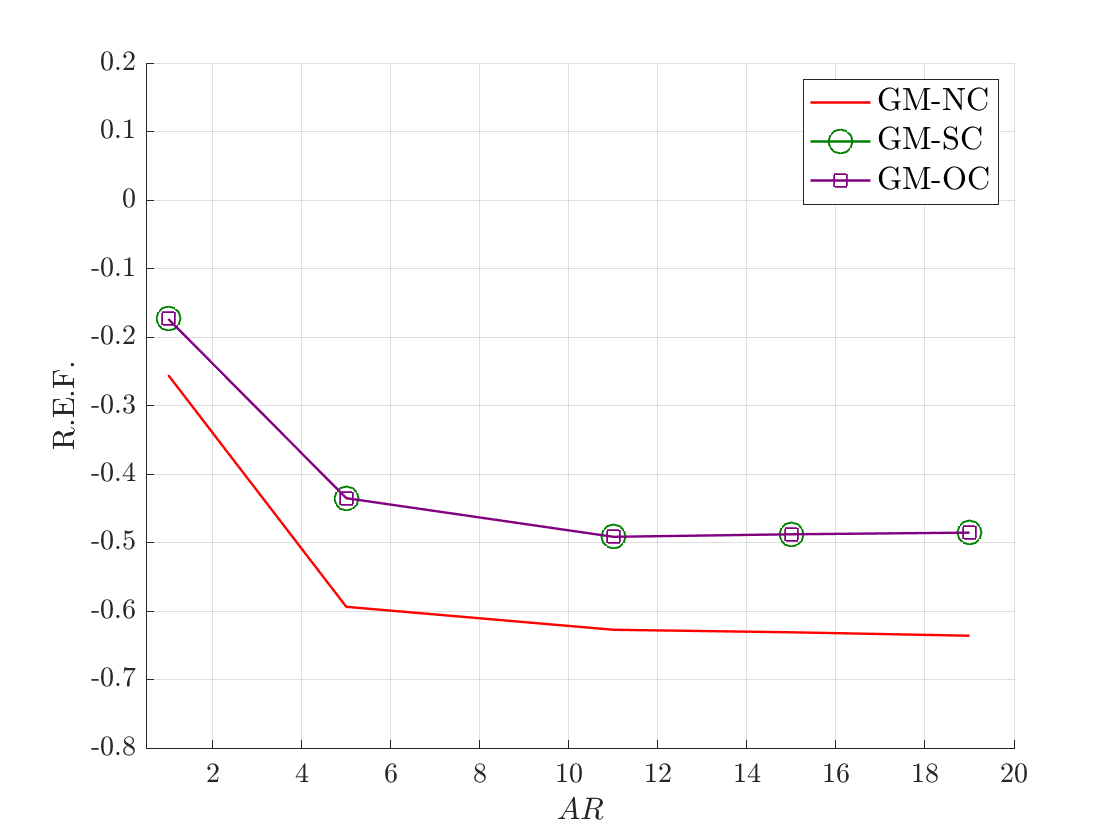}}
    \caption{\textit{A priori} results for anisotropically filtered DNS data based on book-type resolution: (a) Correlation Coefficient (C.C.) and (b) Relative Error in Mean Energy Flux (R.E.F.)}
    \label{fig:Apriori_aniso_book}
\end{figure}

In \figref{Apriori_aniso_book}, we compare the effect of clipping on C.C. and R.E.F. for the book-type anisotropically filtered DNS data based on several filter aspect ratios. We observe that with an increase in the aspect ratio, there is an initial decrease in the correlation with the filtered DNS data. At higher values of the aspect ratio, C.C. asymptotes and there is no further decrease in the correlation. As compared to the case of isotropic filtering, standard clipping has a greater reduction in correlation for anisotropic filtering. We also observe that optimal clipping preserves the correlation coefficient even for anisotropic filtering. Similar trends are also observed for R.E.F., though the use of anisotropic filtering reduces this quantity for all models. The results indicate that increased anisotropy reduces the model dissipation as compared with filtered DNS, which can be partially attributed to higher model backscatter. The use of standard or optimal clipping leads to a higher model dissipation as compared to the gradient model without clipping, and therefore the dissipative performance of these models does not deteriorate as much with increased anisotropy. 

\begin{figure}[ht!]
    \centering
    \subfigure[\label{fig:Apriori_aniso_CC_pencil}]{\includegraphics[width=0.49\textwidth]{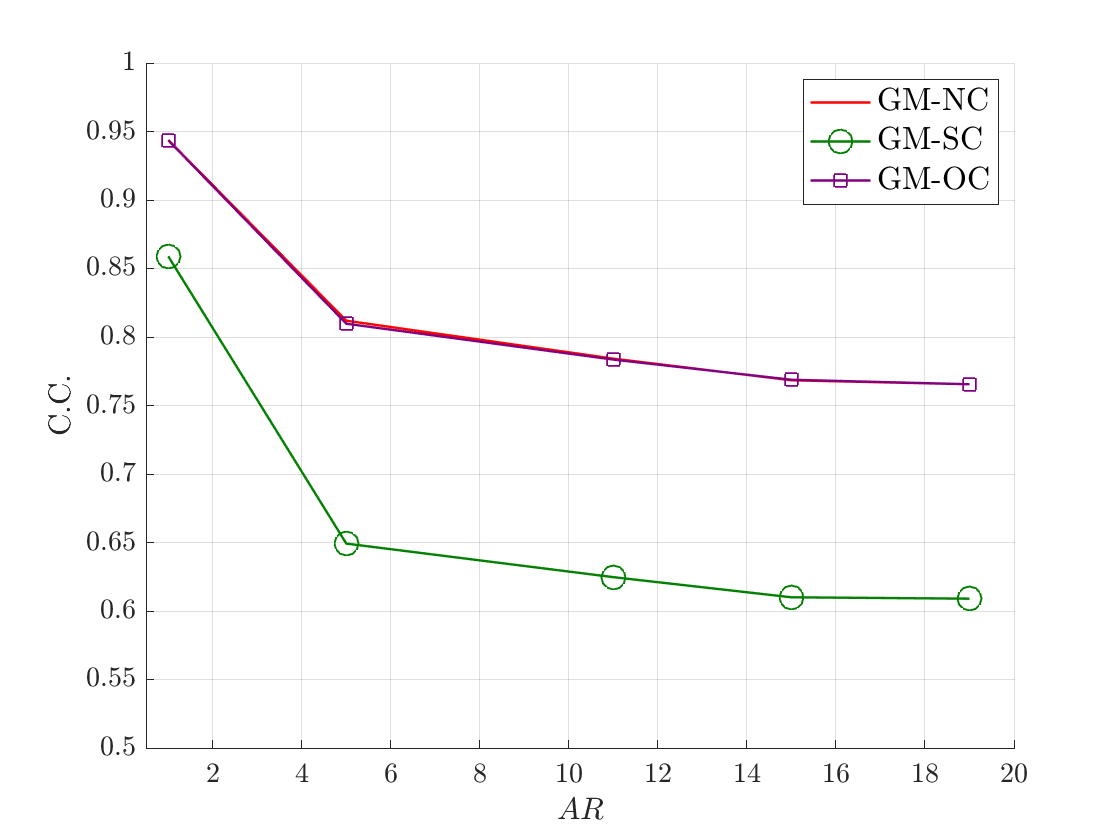}}
    \subfigure[\label{fig:Apriori_aniso_dissip_pencil}]{\includegraphics[width=0.49\textwidth]{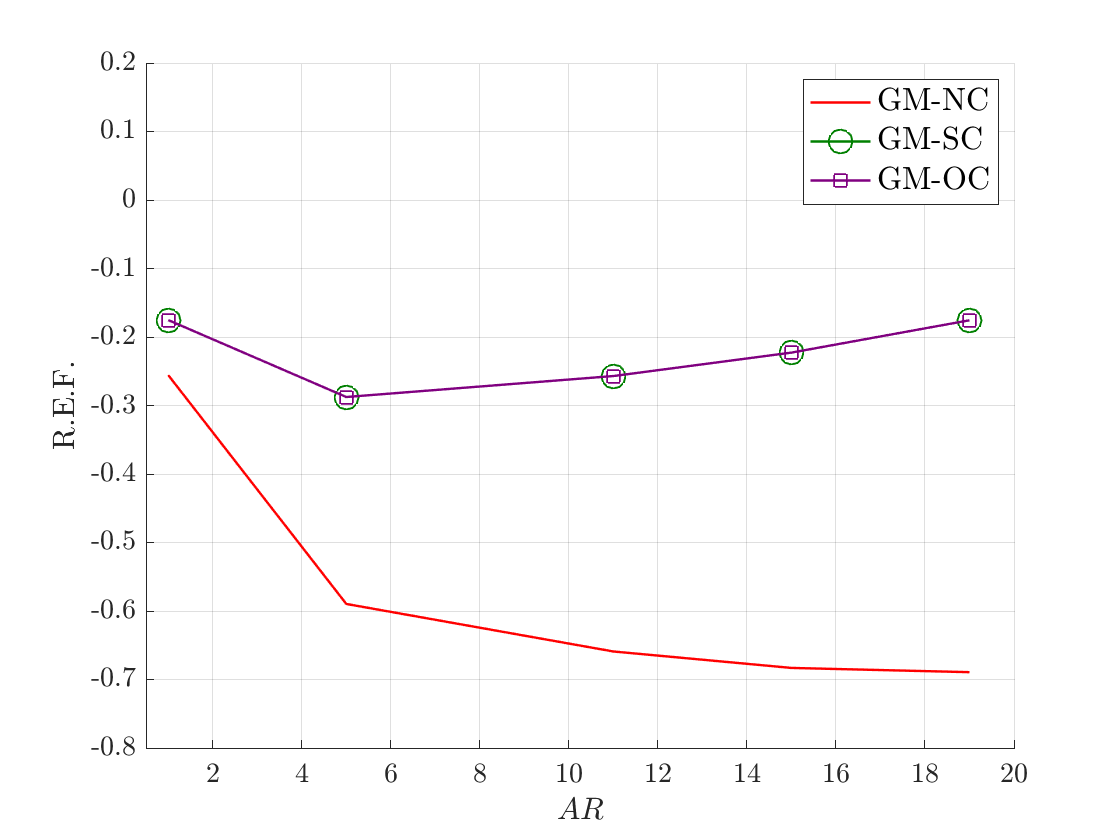}}
    \caption{\textit{A priori} results for anisotropically filtered DNS data based on pencil-type resolution: (a) Correlation Coefficient (C.C.) and (b) Relative Error in Mean Energy Flux (R.E.F.)}
    \label{fig:Apriori_aniso_pencil}
\end{figure}

In \figref{Apriori_aniso_pencil}, we compare the effect of clipping on C.C. and R.E.F. for the pencil-type anisotropically filtered DNS data based on several filter aspect ratios. For this filter shape, we again observe that an increase in the aspect ratio leads to a reduced correlation which seems to asymptote at larger values. The C.C. for standard clipping deteriorates significantly as compared to isotropic filtering. Optimal clipping maintains the correlation of SGS tensor with those from filtered DNS similar to the unclipped gradient model. We also observe that without clipping, the model dissipation decreases significantly at high aspect ratios. Clipping dramatically improves the prediction of SGS dissipation at higher aspect ratios. For all of the above tests, the filter width in the fine direction was kept fixed and the filter widths in the coarse directions were modified to attain the target aspect ratios. However, we have also examined other filter widths in the fine direction and attained similar results.

These results indicate that in the presence of resolution anisotropy, optimal clipping improves the gradient model's prediction of model dissipation while preserving the accuracy of the predicted SGS tensor. These characteristics make the optimally clipped gradient model particularly attractive for LES. It is expected that the improved dissipative characteristics as compared to that of the unclipped gradient model will result in improved numerical stability for higher Reynolds number LES. Moreover, we expect the optimally clipped gradient model's improved structural accuracy as compared to that of standard clipping will result in more accurate LES predictions for anisotropic flows at coarse resolutions. We shall investigate these effects in more detail in the next section.

\subsection{\textit{A posteriori} tests}
\label{sec:Apost}

\textit{A posteriori} simulations were performed using PHASTA \cite{PHASTA} (Parallel Hierarchic Adaptive Stabilized Transient Analysis) which is a SUPG/PSPG/grad-div stabilized finite element based CFD solver \cite{Whiting2001}  that solves the filtered Navier-Stokes equations in advective form. Each simulation uses a piecewise trilinear spatial discretization and the generalized-$\alpha$ method with $\rho_{\infty} = 0.5$ for temporal discretization \cite{Jansen2000}. The code has been validated for LES \citep{Tejada2003,Tejada2005} and DNS \citep{Trofimova2009,Balin2021}. The filter width input for Clark's gradient model is taken to be the same as the computational grid size being used. The dynamic Smagorinsky model, based on the implementation in \cite{Tejada2003}, and the dynamic mixed model use a filter width ratio of $\sqrt{3}$. We follow a similar filtering procedure to \cite{Tejada2003} for the implementation of the dynamic mixed model. 

\subsubsection{Forced HIT}

The first \textit{a posteriori} tests we report on are for forced HIT at a high Reynolds number. For this flow, the turbulence is sustained by an energy cascade in which energy is injected to the largest scales of motion through a prescribed forcing function. Energy is then transferred to successively smaller scales through inertial processes, and energy is finally dissipated at the smallest scales through viscous interactions. In our calculations, the domain is taken to be a tri-periodic box of side length $2 \pi$, and following \cite{Bazilevs2007}, the prescribed forcing is taken to be:

\begin{equation}
    \bar{\pmb{f}} (\pmb{x}) = \sum_{\substack{\pmb{k} \\ \vert k_i \vert < k_f \\ \pmb{k} \neq 0}} \frac{P_{in}}{2 E_{kf}} \hat{\pmb{u}}_k \exp{(\iota \pmb{k}\cdot\pmb{x})},
\end{equation}

\begin{equation}
    E_{kf} = \frac{1}{2} \sum_{\substack{\pmb{k} \\ \vert k_i \vert < k_f \\ \pmb{k} \neq 0}} \hat{\pmb{u}}_k \cdot \hat{\pmb{u}}_k  ,
\end{equation}

\begin{equation}
    \hat{\pmb{u}}_k = \frac{1}{\vert \Omega \vert} \int_{\Omega} \pmb{u}^h (\pmb{x}) \exp{(-\iota \pmb{k}\cdot\pmb{x})},
\end{equation}

\noindent where $P_{in}$ is the power input to the forcing and $k_f$ is the wavenumber magnitude to be forced. The values of these quantities are chosen as 62.843 and 3, respectively, to be consistent with \cite{Bazilevs2007}. To simulate high Reynolds number for the flow, we set the viscosity to a very small value, namely $10^{-12}$. This ensures that the energy dissipation due to viscous effects is dominated by energy dissipation due to the SGS model and numerical discretization. Furthermore, numerical experiments show that for this flow the numerical dissipation from SUPG/PSPG/grad-div stabilization is significantly smaller in comparison to the SGS model dissipation for each considered explicit SGS model. Therefore, the SGS model serves as the dominant mode of energy dissipation. This enables us to amplify the importance of clipping whose effect is obscured for flows at low $Re_{\lambda}$. At such a high value of $Re_{\lambda}$, DNS results are not available for comparison. We instead compare LES energy spectra to the theoretical Kolmogorov spectrum \cite{Pope2000}:

\begin{equation}
    E(k) = C \epsilon^{2/3} k^{-5/3},
\end{equation}

\noindent where the constant $C$ is taken to be $1.6$ following experimental and theoretical observations \cite{Pope2000}. The value of dissipation ($\epsilon$) is equal to the power input ($P_{in}$) as the turbulence is in equilibrium. In the figures, K41 refers to the theoretical results.  We discretize the domain with three mesh resolutions composed of 32, 64 and 128 uniformly spaced elements in each direction. For the SGS model, the filter width is selected to be $2\pi/32$, $2\pi/64$ and $2\pi/128$ for meshes with $32^3$, $64^3$ and $128^3$ elements respectively. 

\begin{figure}
    \centering
    \subfigure[\label{NM_HIT}]{\includegraphics[width=0.49\textwidth]{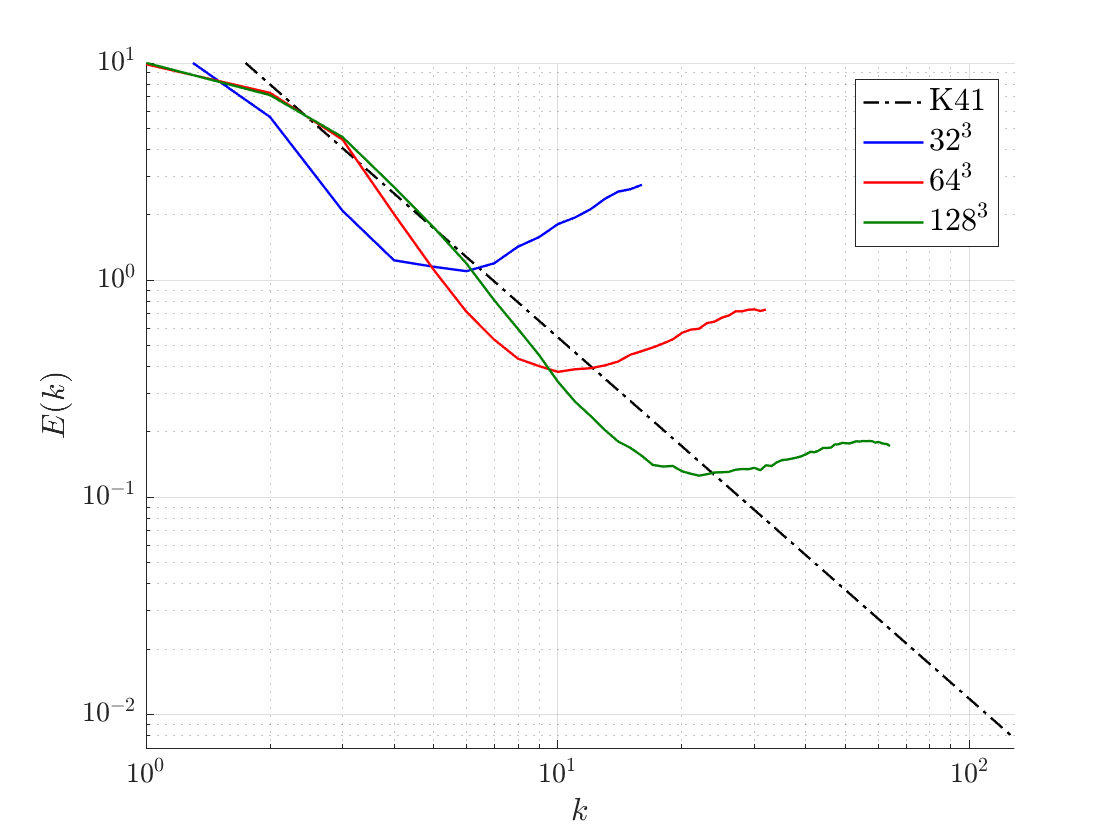}}
    \subfigure[\label{DS_HIT}]{\includegraphics[width=0.49\textwidth]{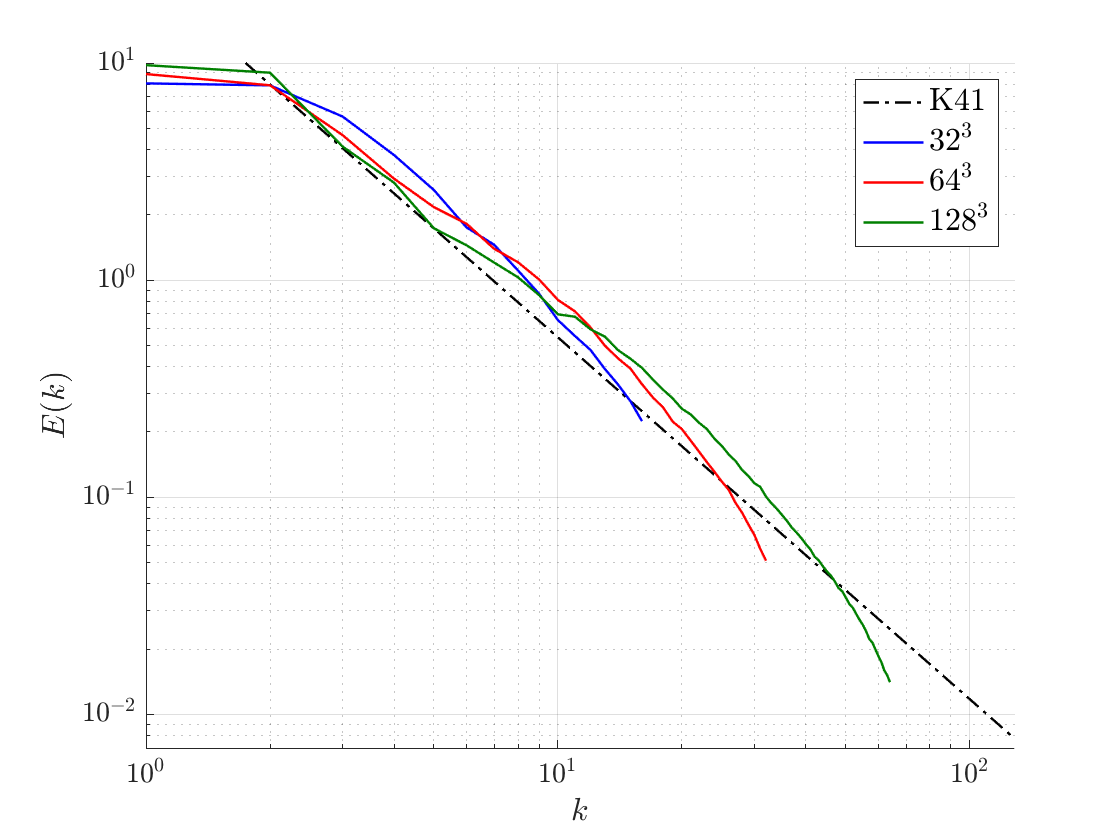}}
    \subfigure[\label{GM_NC_HIT}]{\includegraphics[width=0.49\textwidth]{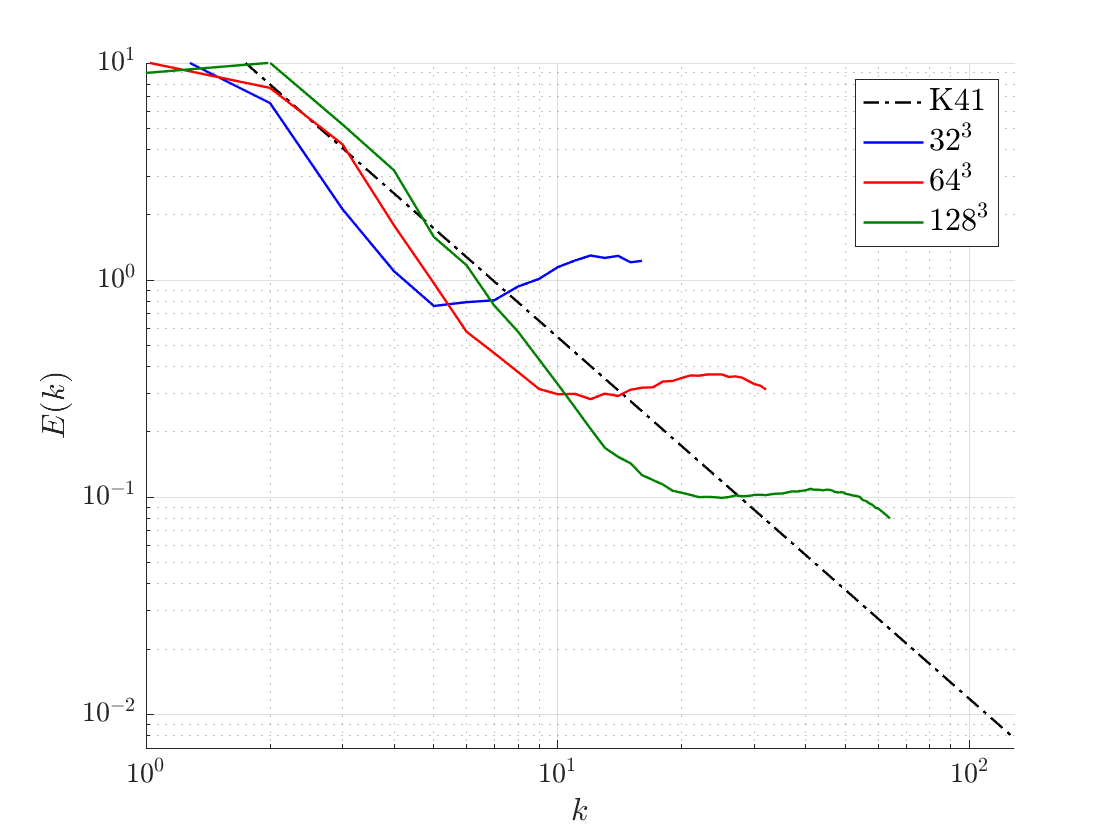}}
    \subfigure[\label{GM_SC_HIT}]{\includegraphics[width=0.49\textwidth]{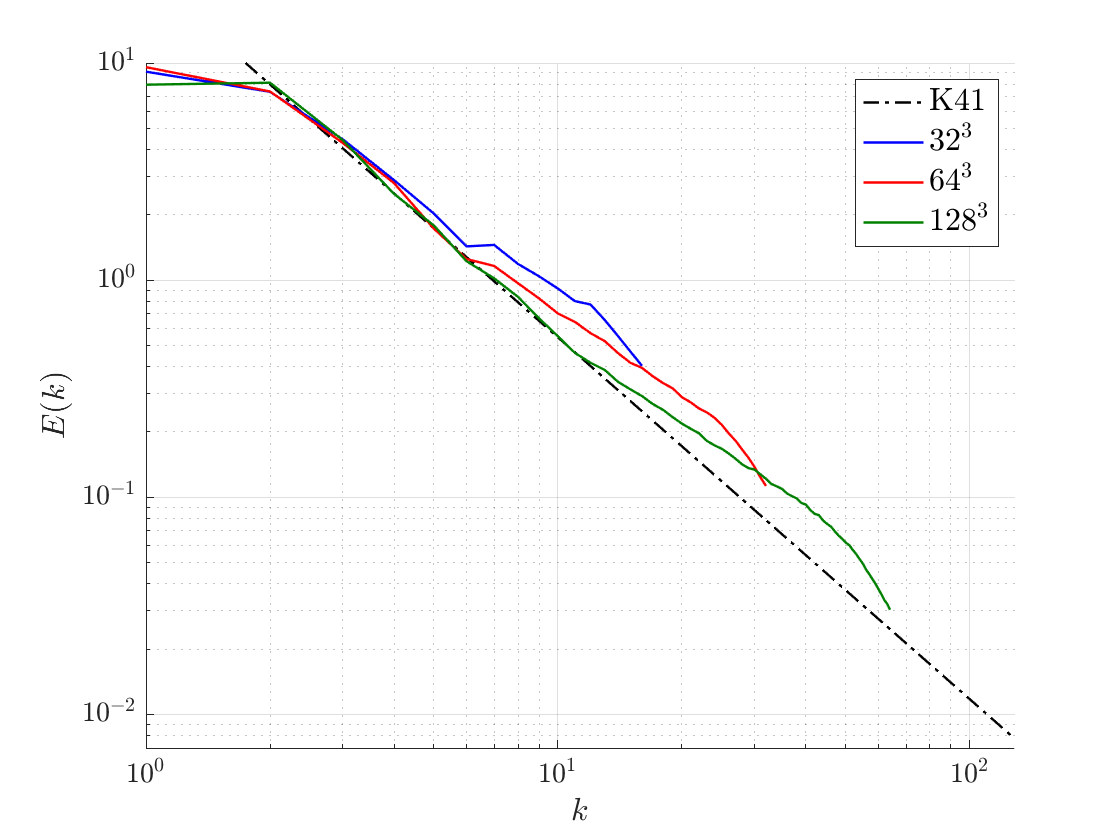}}
    \subfigure[\label{GM_DMM_HIT}]{\includegraphics[width=0.49\textwidth]{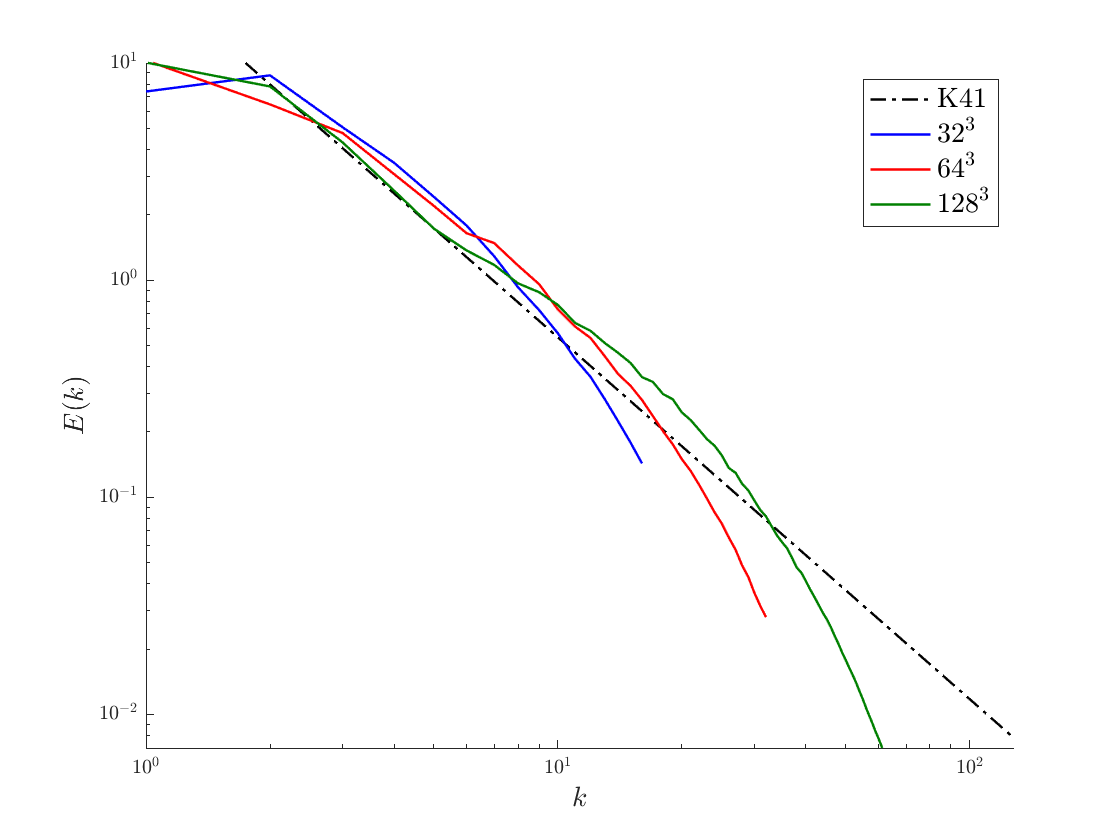}}
    \subfigure[\label{GM_OC_HIT}]{\includegraphics[width=0.49\textwidth]{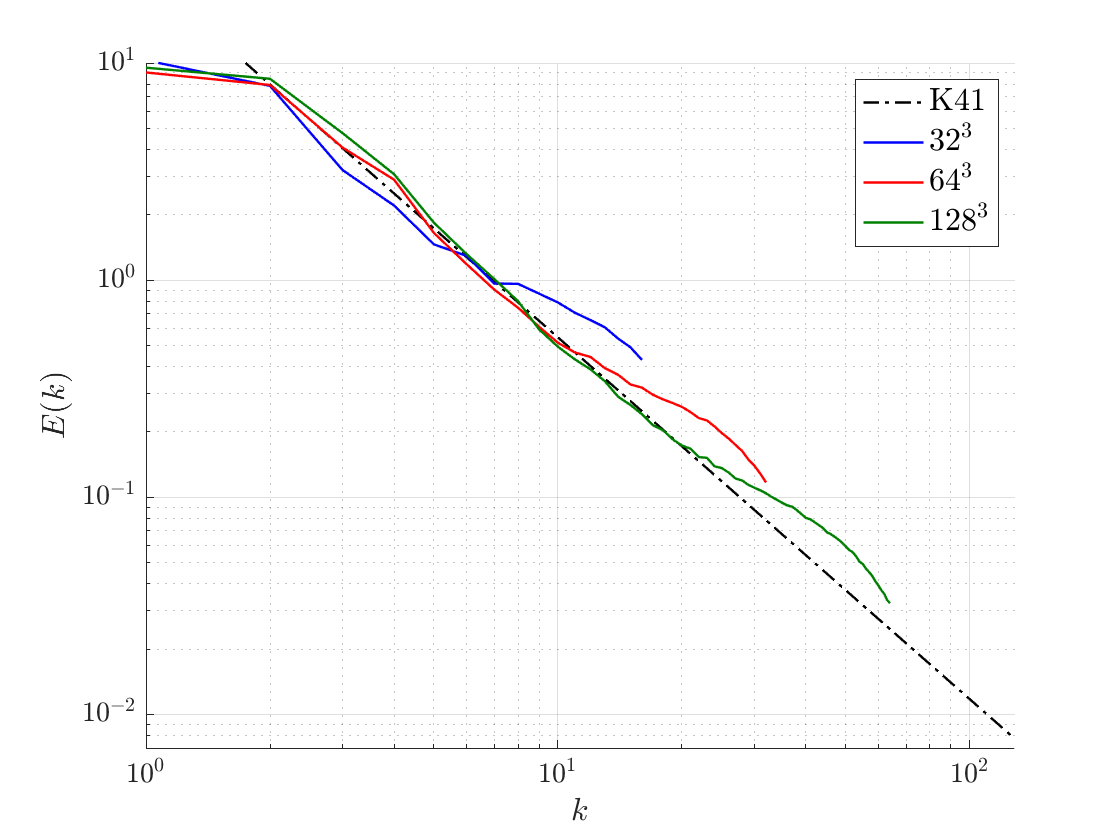}}
    
    \caption{Energy spectra for HIT at a high Reynolds number as predicted using (a) no model, (b) the dynamic Smagorinsky model, (c) the gradient model without clipping, (d) the gradient model with standard clipping, (e) the dynamic mixed model and (f) the gradient model with optimal clipping}
    \label{fig:ReInf}
\end{figure}

The energy spectra obtained using various SGS models are shown in \figref{ReInf}. When no explicit SGS model is used, we observe a large pileup of energy at the highest resolved wavenumbers. This result indicates that the numerical dissipation by itself is insufficient to adequately dissipate the energy cascading down to the unresolved scales from the resolved scales. The dynamic Smagorinsky model does not lead to a pileup of energy and obeys the $5/3^{\text{rd}}$ slope of the theoretical spectra. However, it does exhibit an overprediction of energy at intermediate wavenumbers between the lowest and highest resolved wavenumbers. A large pileup of energy is also exhibited by the gradient model without clipping. This result indicates that the gradient model by itself does not provide sufficient model dissipation. We observe that standard and optimal clipping yield significantly improved energy spectra predictions. Optimal clipping leads to a slightly longer inertial subrange and less energy pileup. Both the gradient model with standard and the gradient model with optimal clipping also provide better predictions at intermediate wavenumbers as compared to the dynamic Smagorinsky model. The dynamic mixed model also does not exhibit a pileup of energy but it overpredicts energy at the intermediate wavenumbers. The results indicate that at higher Reynolds numbers the original model form is insufficient and requires additional mechanisms such as standard clipping, optimal clipping, or the inclusion of a Smagorinsky model like term to avoid energy pileup at the highest resolved wavenumbers. As we demonstrated that using no explicit SGS model is insufficient at high Reynolds numbers for forced HIT, we will only consider explicit SGS models for next flow cases.

\subsubsection{3-D Taylor Green Vortex flow at $Re = 1600$}

The next \textit{a posteriori} tests we consider are for 3-D Taylor Green Vortex (TGV) flow. For sufficiently high $Re$, TGV flow starts as laminar, then transitions to fully turbulent flow, and after the flow fully transitions, the turbulence decays. Our calculations use a tri-periodic cubical domain of $2\pi$ side length. The flow is started from the initial laminar velocity profile:

\begin{equation}
    \textbf{u} = \begin{bmatrix} \sin{(x_1)} \cos{(x_2)} \cos{(x_3)} \\ - \cos{(x_1)} \sin{(x_2)} \cos{(x_3)} \\ 0
    \end{bmatrix},
\end{equation}

\noindent and no external forcing is applied. We consider a Reynolds number of $Re = 1600$, where $Re = U L / \nu$ and $U = L = 1$. In this paper, the temporal evolution of resolved kinetic energy and dissipation are investigated. The definitions of these quantities are given in \eref{tke_tg} and \eref{dissip_tg} below:

\begin{equation}
    E_k (t) = \frac{1}{\vert \Omega \vert} \int_{\Omega} \frac{\pmb{u}(t) \cdot \pmb{u}(t)}{2} d\Omega
    \label{eq:tke_tg}
\end{equation}

\begin{equation}
    \epsilon (t) = 2 \nu \frac{1}{\vert \Omega \vert} \int_{\Omega} \frac{\pmb{\omega} (t) \cdot \pmb{\omega}(t)}{2} d\Omega
    \label{eq:dissip_tg}
\end{equation}

\noindent where $\pmb{\omega}$ is vorticity. We evaluate these statistics for a coarse mesh with $64^3$ elements and a fine mesh with $128^3$ elements, which correspond to filter widths of $2\pi/64$ and $2\pi/128$ respectively. These quantities are compared with DNS and filtered DNS results attained from \cite{Shoraka2017}. In this paper, we only examine the temporal evolution of resolved turbulent kinetic energy and dissipation until a fully turbulent state is reached. The decay of turbulence depends on the initial conditions at the start of this regime. As different models yield different flow states at the onset of turbulence decay, it is difficult to determine whether model deficiencies in the transition or decay regions are responsible for poor performance as the turbulence decays.

\begin{figure}[b!]
    \centering
    \subfigure[\label{Ek_64}]{\includegraphics[width=0.49\textwidth]{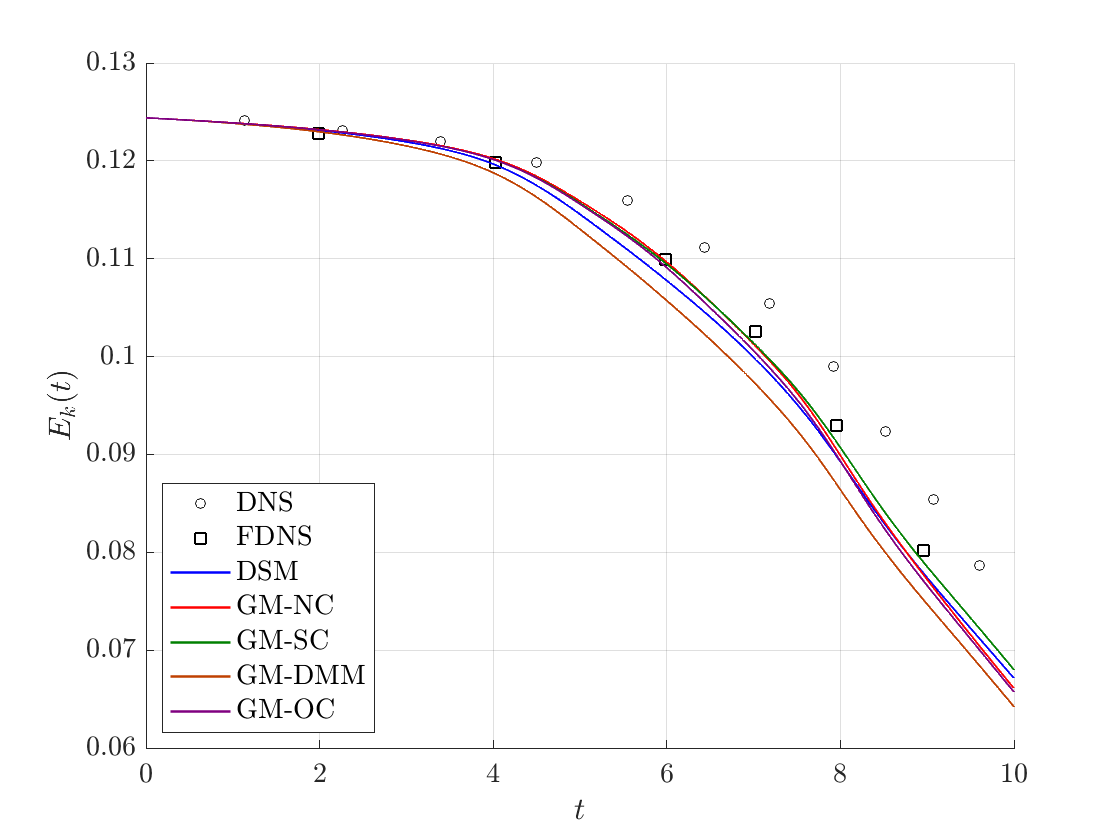}}
    \subfigure[\label{Ek_128}]{\includegraphics[width=0.49\textwidth]{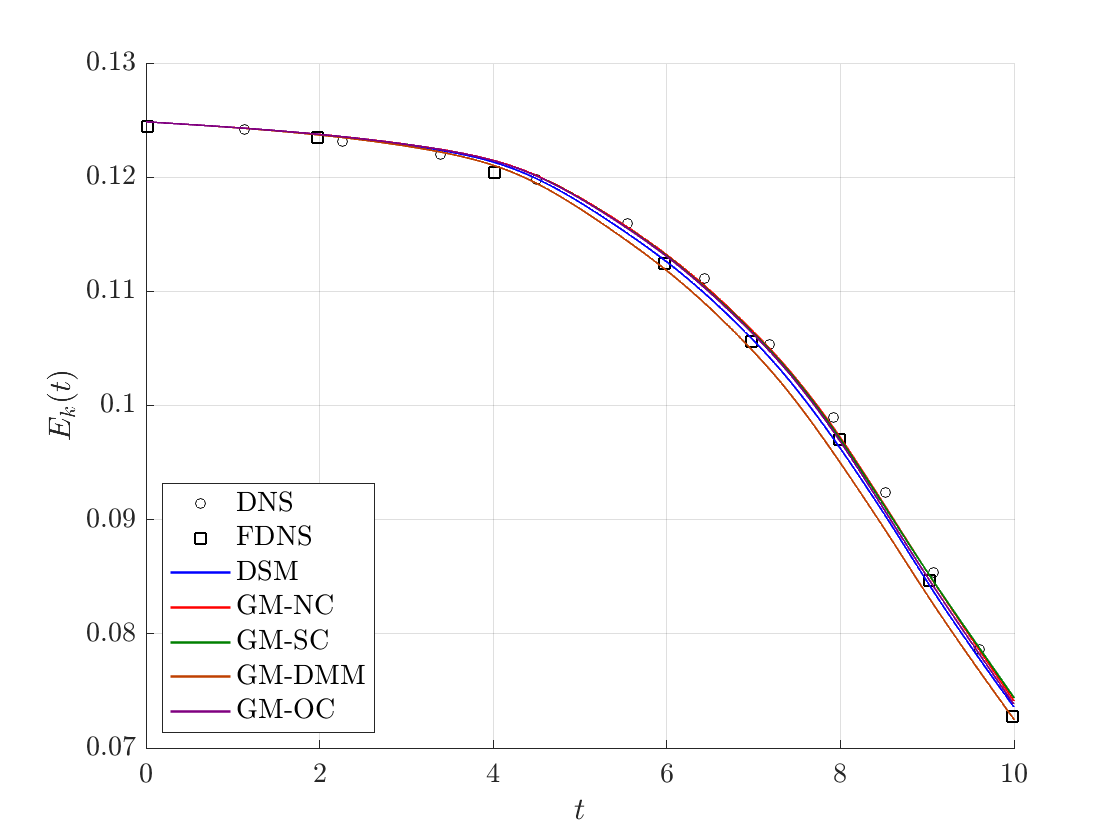}}
    \caption{Resolved kinetic energy ($E_k$) for TGV flow at $Re = 1600$ for (a) $64^3$ and (b) $128^3$ element meshes}
    \label{fig:Ek_TGV}
\end{figure}

\figref{Ek_TGV} shows the evolution of resolved kinetic energy for the $64^3$ and $128^3$ element meshes. For all SGS models, we expect the results to match with the filtered DNS results as we refine the mesh. This is indeed the case for the fine mesh, and all models have similar evolution of resolved kinetic energy for this mesh. All models except the dynamic mixed model also yield accurate predictions of the filtered DNS turbulent kinetic energy for the coarse mesh resolution. The dynamic mixed model underpredicts the filtered DNS turbulent kinetic energy for the coarse mesh, highlighting that this model exhibits a faster temporal decay in resolved kinetic energy compared with the other models due to its overdissipative nature. 

\begin{figure}[t!]
    \centering
    \subfigure[\label{Dissip_64}]{\includegraphics[width=0.49\textwidth]{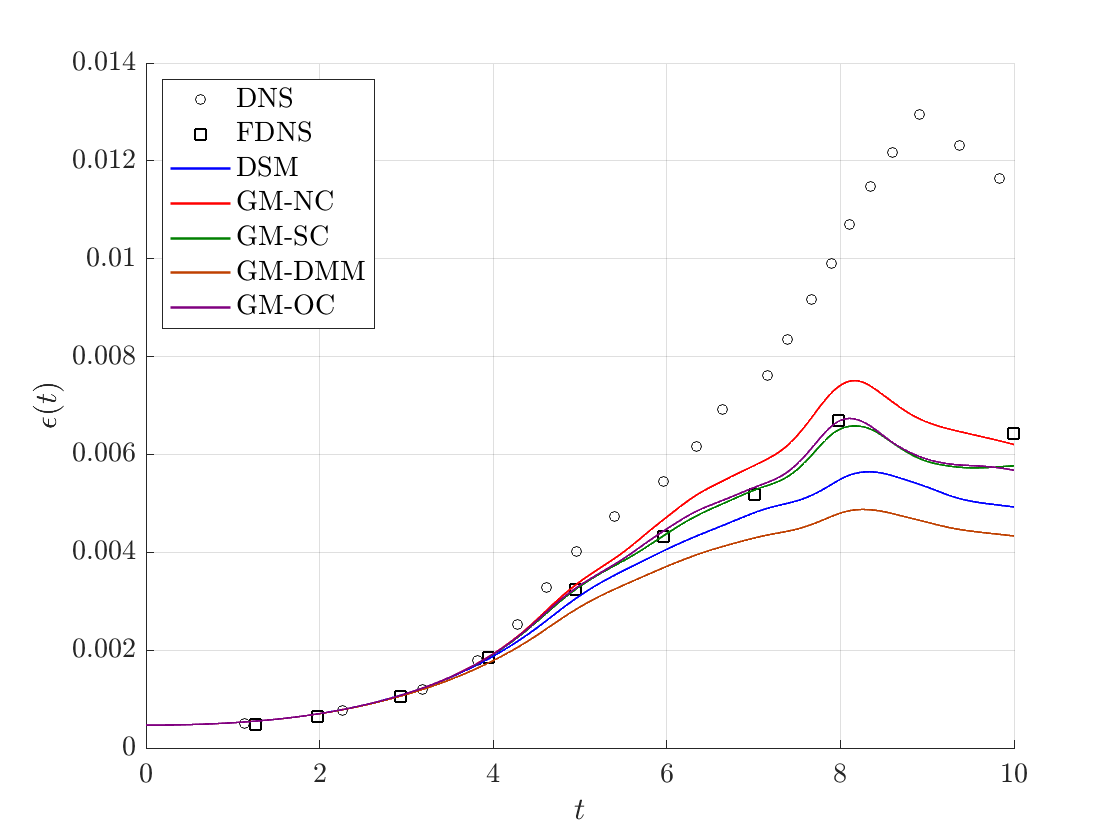}}
    \subfigure[\label{Dissip_128}]{\includegraphics[width=0.49\textwidth]{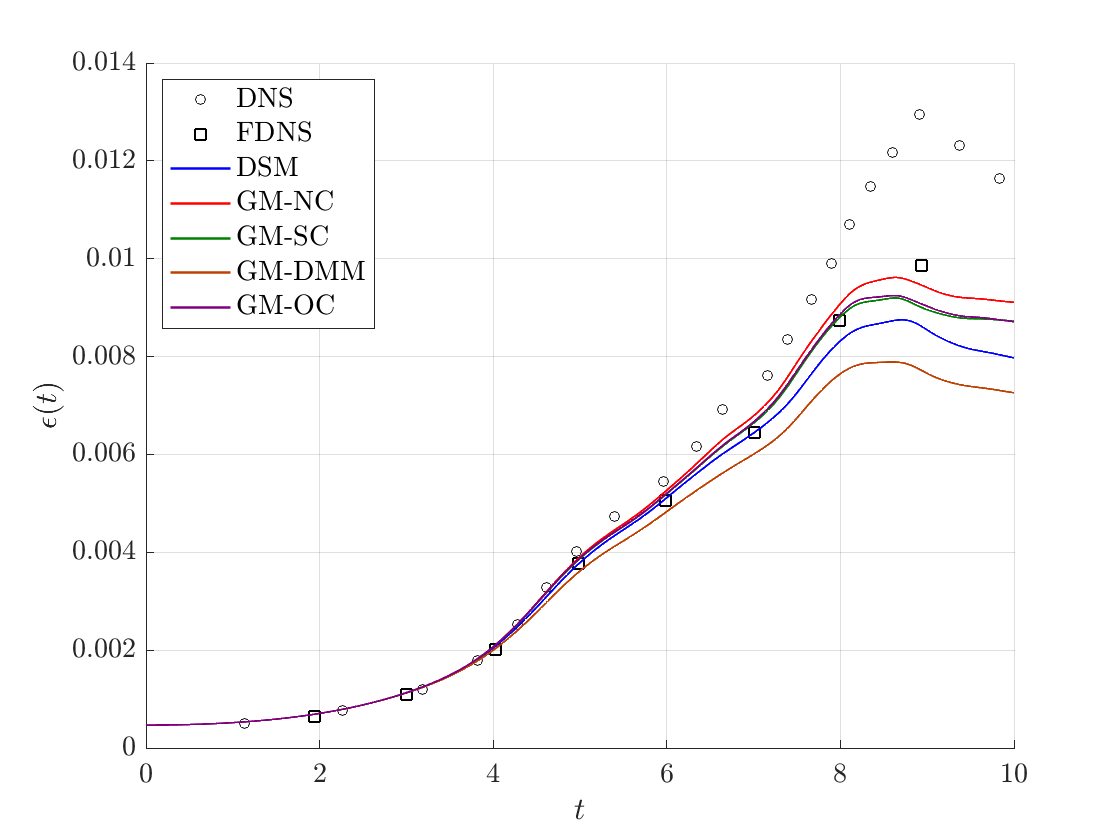}}
    \caption{Resolved dissipation for TGV flow at $Re = 1600$ for (a) $64^3$ and (b) $128^3$ element meshes}
    \label{fig:Dissip_TGV}
\end{figure}

\figref{Dissip_TGV} displays the temporal evolution of the resolved dissipation for the two mesh resolutions. The dynamic Smagorinsky model underpredicts the resolved dissipation computed using the filtered DNS data. The dynamic mixed model underpredicts the resolved dissipation to an even greater degree, and even for the fine mesh resolution, the underprediction of resolved dissipation is significant. The gradient model without clipping overpredicts the resolved dissipation, especially for the coarse mesh resolution. With standard and optimal clipping, we attain a more accurate prediction of resolved dissipation as compared to the other models. 

\begin{figure}[t!]
    \centering
    \subfigure[\label{ES_64}]{\includegraphics[width=0.49\textwidth]{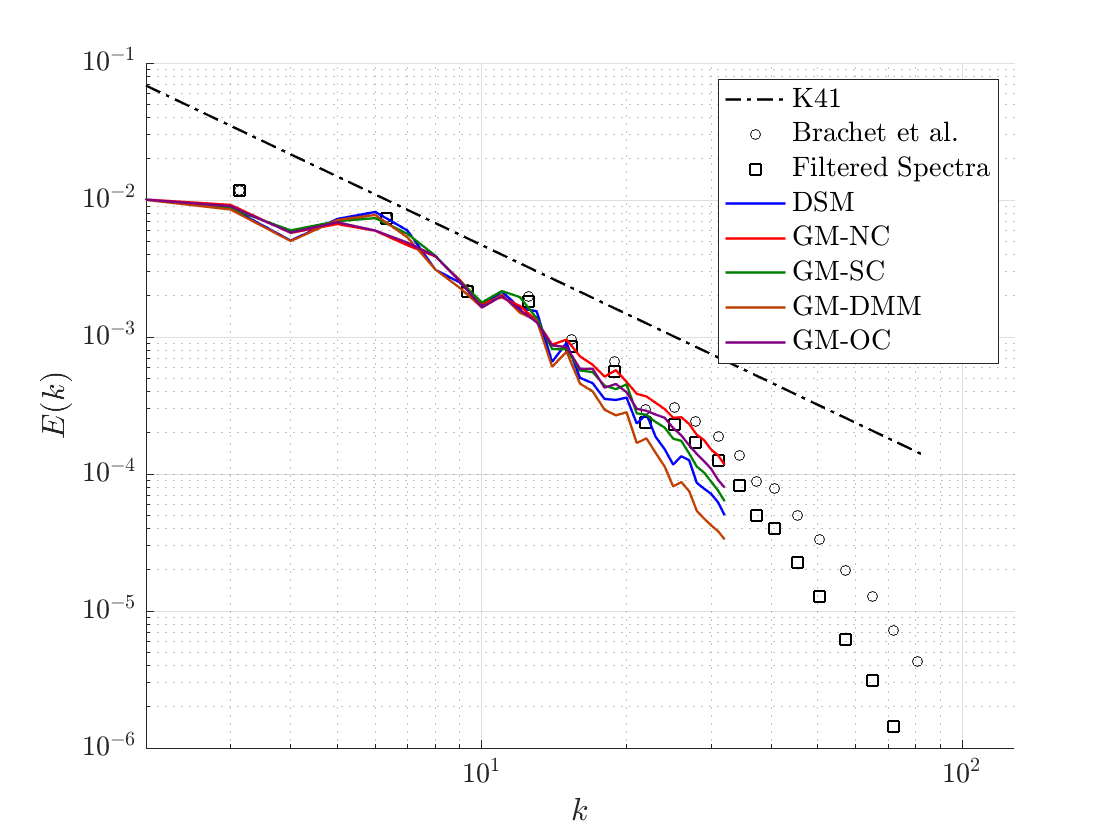}}
    \subfigure[\label{ES_128}]{\includegraphics[width=0.49\textwidth]{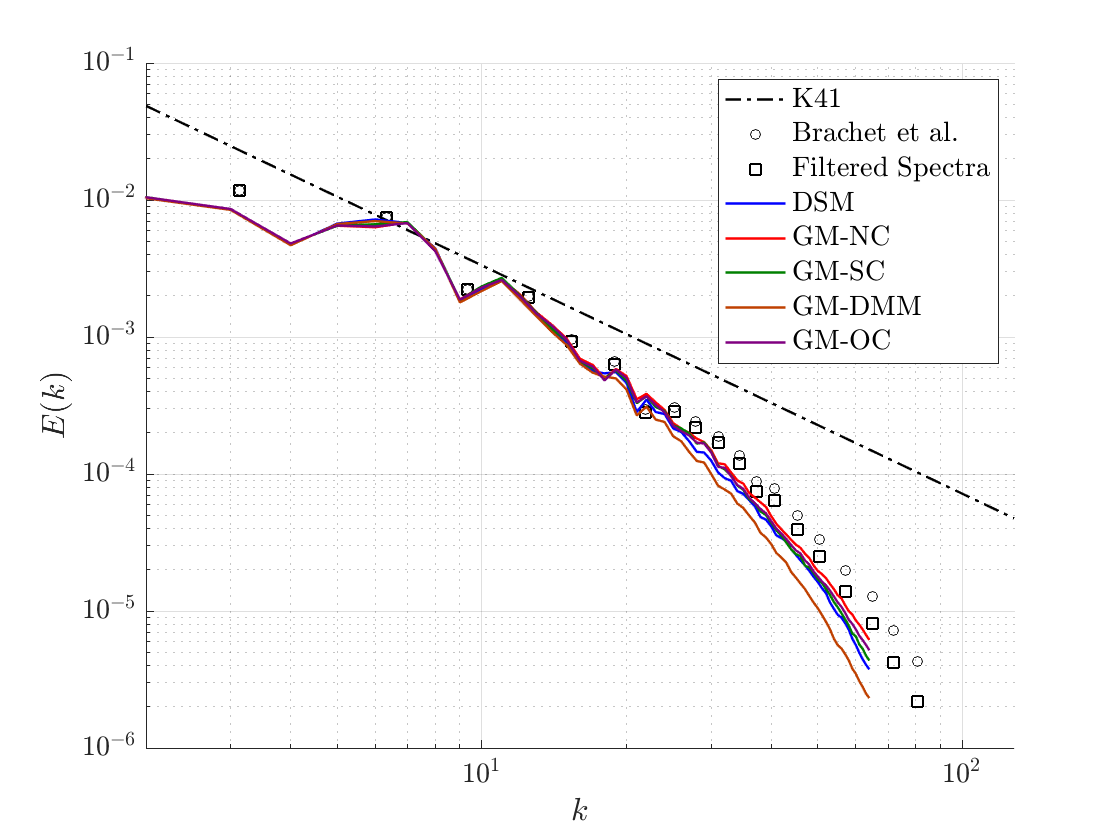}}
    \caption{Energy spectra at $t = 9$ for TGV flow at $Re = 1600$ for (a) $64^3$ and (b) $128^3$ element meshes}
    \label{fig:ES_TGV}
\end{figure}

In \figref{ES_TGV}, we compare the energy spectra at $t = 9$ attained using the SGS models to that attained using DNS \cite{Brachet1983} and filtered DNS results. For the filtered DNS spectra, we apply box-filter-based differential filtering \cite{Germano1986} to the DNS results. The dynamic Smagorinsky and dynamic mixed models underpredict the energy spectra at higher wavenumbers, with the latter model leading to a more severe underprediction. This indicates that the dynamic mixed model has too much model dissipation that is leading to a reduction in turbulent kinetic energy content at the smallest resolved scales. On the other hand, the energy spectra predictions by the gradient model without clipping, standard clipping, and optimal clipping are closest to the filtered DNS spectra. For the coarse mesh resolution, optimal clipping gives slightly better results than standard clipping. From this test case, we conclude that, for the gradient model, standard and optimal clipping exhibit the best performance of all considered SGS models for the estimation of considered quantities of interest for laminar-turbulent transition in TGV flow at $Re = 1600$. Furthermore, the distribution of energy among the wavenumbers is also accurately captured by these models, with optimal clipping performing slightly better than standard clipping.

\subsubsection{Turbulent channel flow at $Re_{\tau} = 590$}

As a final test case, we investigate the performance of optimal clipping in the presence of walls. In particular, we consider the classical problem of turbulent channel flow at $Re_{\tau} = 590$ and compare mean flow and stress profiles with those from DNS \cite{Moser1999}. A domain of $2 \pi \delta \; \times 2 \delta \times \pi \delta$, where $\delta$ ($ = 1 $) is the channel half-height, is employed in our calculations. Periodic boundary conditions are applied in the spanwise and streamwise directions. No-slip boundary conditions are imposed at the walls, that is at $y=0$ and $y=2 \delta$. The flow is initialized with a velocity profile as per the law of the wall with superimposed random perturbations. The flow is driven by a fixed-mass forcing which is based on the mass flux at the inflow corresponding to a bulk Reynolds number ($Re_b = \bar{u}_b \delta / \nu$ where $\bar{u}_b$ is the bulk velocity) of  $10975$. After a statistically stationary state is reached, the spanwise and streamwise averaged flow variables are time-averaged over an interval of at least $30T$, where $T = \bar{u}_b/L_x$ is the flow-through time, and $L_x$ is the streamwise length of the domain. Split-window time averaging analysis of velocity and stress profiles shows that the statistics are converged by this time. We consider two mesh resolutions and these details are tabulated in \tabref{Channel_mesh}. In \tabref{Channel_mesh}, $\Delta x^+$ is the streamwise spacing, $\Delta z^+$ is the spanwise spacing, $\Delta y_1^+$ is the wall-normal spacing for the first element off-the-wall, and $\Delta y_c^+$ is the wall-normal spacing at the center of the channel, all non-dimensionalized in inner-layer units.

\begin{table}[b!]
    \centering
    \begin{tabular}{ccccccc}
        \hline
        \hline
         $Re_{\tau}$ & \textbf{Mesh Resolution} & \textbf{Number of Elements} & \textbf{$\Delta x^+$} & \textbf{$\Delta y_1^+$} & \textbf{$\Delta y_c^+$} & \textbf{$\Delta z^+$} \\
         \hline
         590 & Coarse & 48 $\times$ 111 $\times$ 48 & 78 & 1 & 45 & 39 \\ 
         590 & Medium & 64 $\times$ 133 $\times$ 64 & 58 & 1 & 36 & 29 \\
         \hline
    \end{tabular}
    \caption{Mesh parameters for the turbulent channel flow case}
    \label{tab:Channel_mesh}
\end{table}


\begin{table}[b!]
    \centering
    \begin{tabular}{ccccccc}
        \hline
        \hline
         \textbf{Mesh Resolution} & \textbf{DNS} & \textbf{DS} & \textbf{AGM-NC} & \textbf{AGM-SC} & \textbf{AGM-DMM} & \textbf{AGM-OC} \\
         \hline
         Coarse & 0.0058 & 0.0047 & 0.0063 & 0.0047 & 0.0053 & 0.0051 \\   
         Medium & 0.0058 & 0.0049 & 0.0063 & 0.0050 & 0.0055 & 0.0053 \\ 
         \hline
    \end{tabular}
    \caption{Skin-friction coefficient prediction for turbulent channel flow at $Re_{\tau} = 590$}
    \label{tab:Cf_ReT590}
\end{table}

Skin friction coefficient predictions, $C_f = \tau_w/( \frac{1}{2} \rho\bar{u}_b^2)$, for both coarse and medium mesh resolutions are shown in \tabref{Cf_ReT590}. As we use fixed-mass forcing, $\bar{u}_{\tau}$ predictions differ between various SGS models yielding differences in predicted skin-friction coefficients. We observe that the dynamic Smagorinsky model and the gradient model with standard clipping significantly under-predict the skin-friction coefficient compared to DNS values for both mesh resolutions. On the other hand, the gradient model without clipping overpredicts the skin-friction coefficient. The gradient model with optimal clipping and the dynamic mixed model have closer predictions to the DNS compared to standard clipping, with the dynamic mixed model yielding the best results. As $\bar{u}_{\tau}$ predictions differ between the considered SGS models, we choose bulk velocity for non-dimensionalizing the velocity and stress profiles.

\begin{figure}[b!]
    \centering
    \subfigure[\label{fig:590_up_coarse}]{\includegraphics[width=0.49\textwidth]{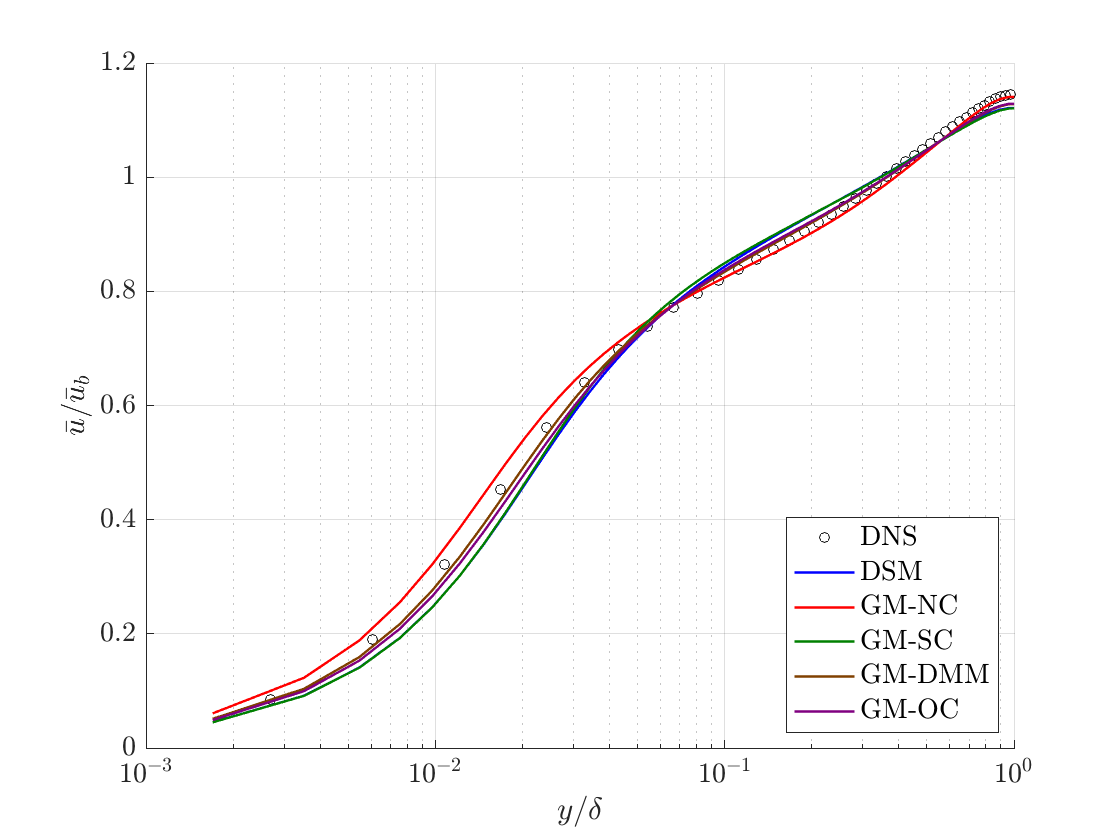}}
    \subfigure[\label{fig:590_up_medium}]{\includegraphics[width=0.49\textwidth]{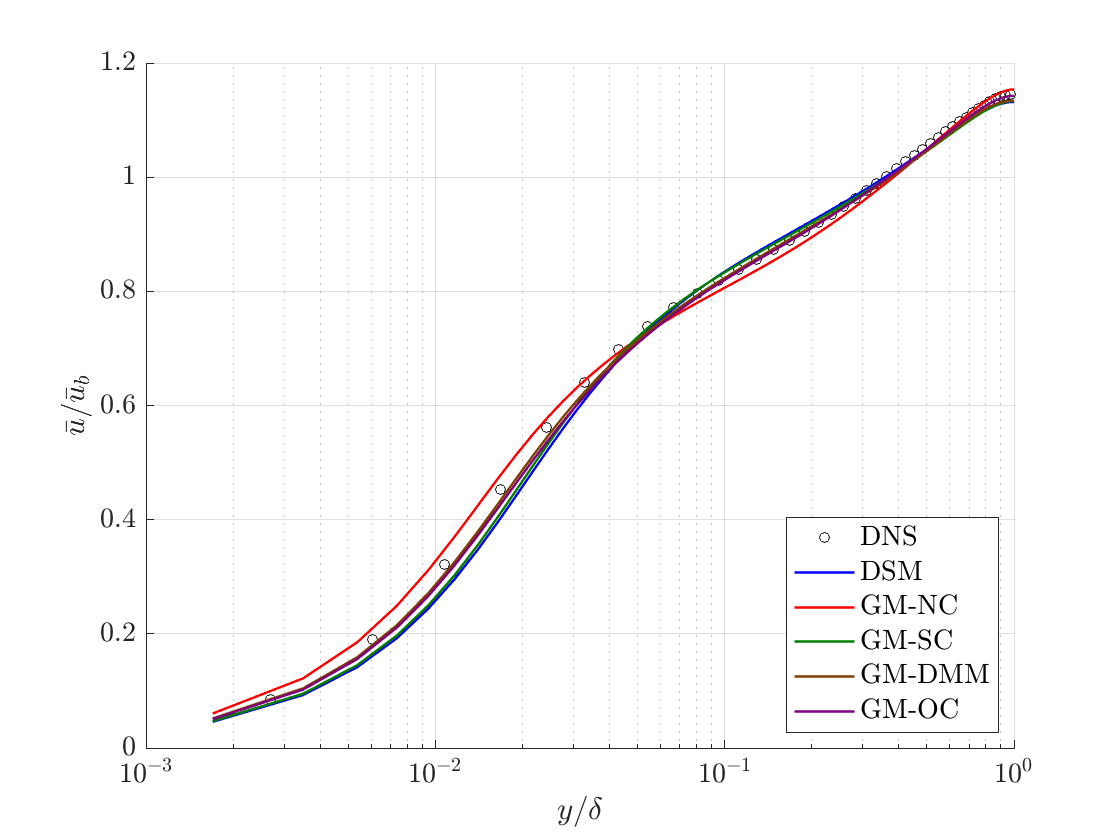}}    
    \caption{Streamwise mean velocity profile for the (a) coarse mesh case and (b) medium mesh for turbulent channel flow at $Re_{\tau} = 590$}
    \label{fig:Channel_590_up}
\end{figure}


Streamwise mean velocity predictions are shown in \figref{Channel_590_up}. We observe that both the dynamic Smagorinsky model and the gradient model with standard clipping underpredict the DNS velocity profile close to the wall and overpredict the velocity profile in the log-linear region. On the other hand, the gradient model without clipping leads to an overprediction of the DNS velocity profile close to the wall. The gradient model with optimal clipping and the dynamic mixed model give more accurate predictions than those attained by the other explicit SGS models. The predictions by these two models are close to each other with the dynamic mixed model predictions being slightly more accurate for the coarse mesh resolution.

\begin{figure}[ht!]
    \centering
    \subfigure[\label{fig:590_k_coarse}]{\includegraphics[width=0.49\textwidth]{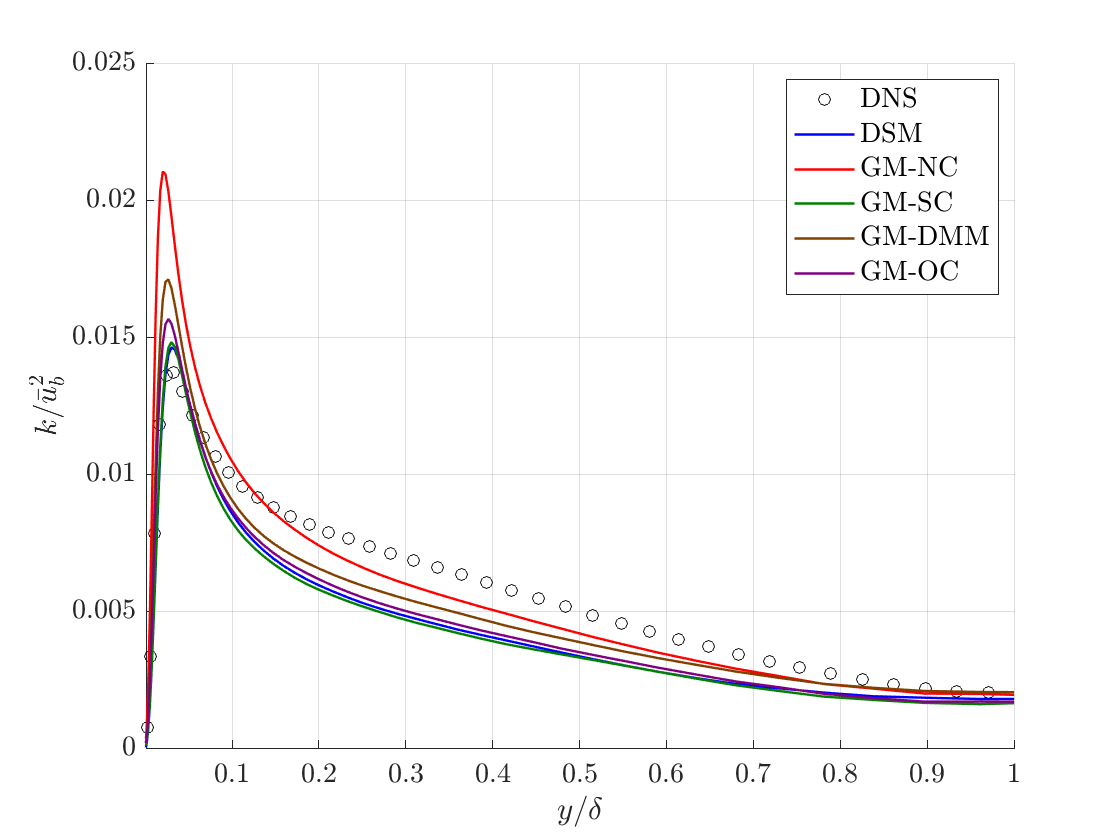}}
    \subfigure[\label{fig:590_a12_coarse}]{\includegraphics[width=0.49\textwidth]{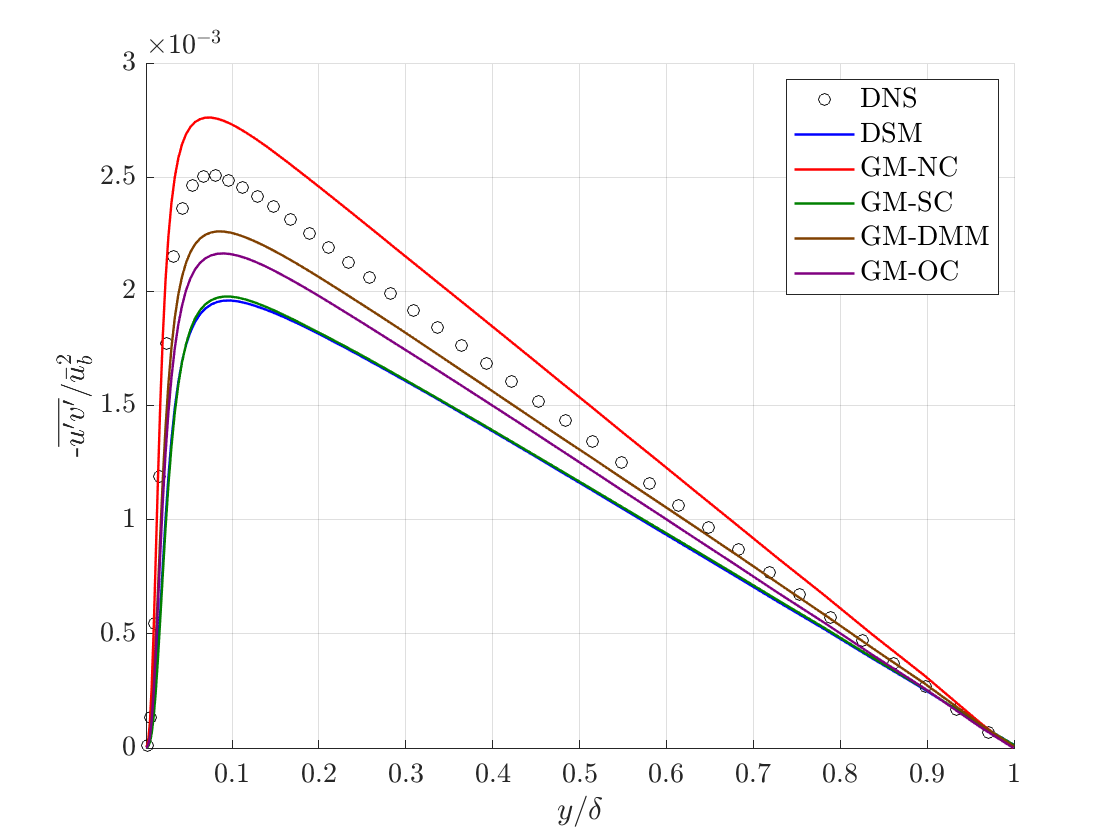}}
    \caption{ (a) Turbulent kinetic energy and (b) Reynolds shear stress for the coarse mesh case for turbulent channel flow at $Re_{\tau} = 590$}
    \label{fig:Channel_590_coarse}
\end{figure}

\begin{figure}[ht!]
    \centering
    \subfigure[\label{fig:590_k_medium}]{\includegraphics[width=0.49\textwidth]{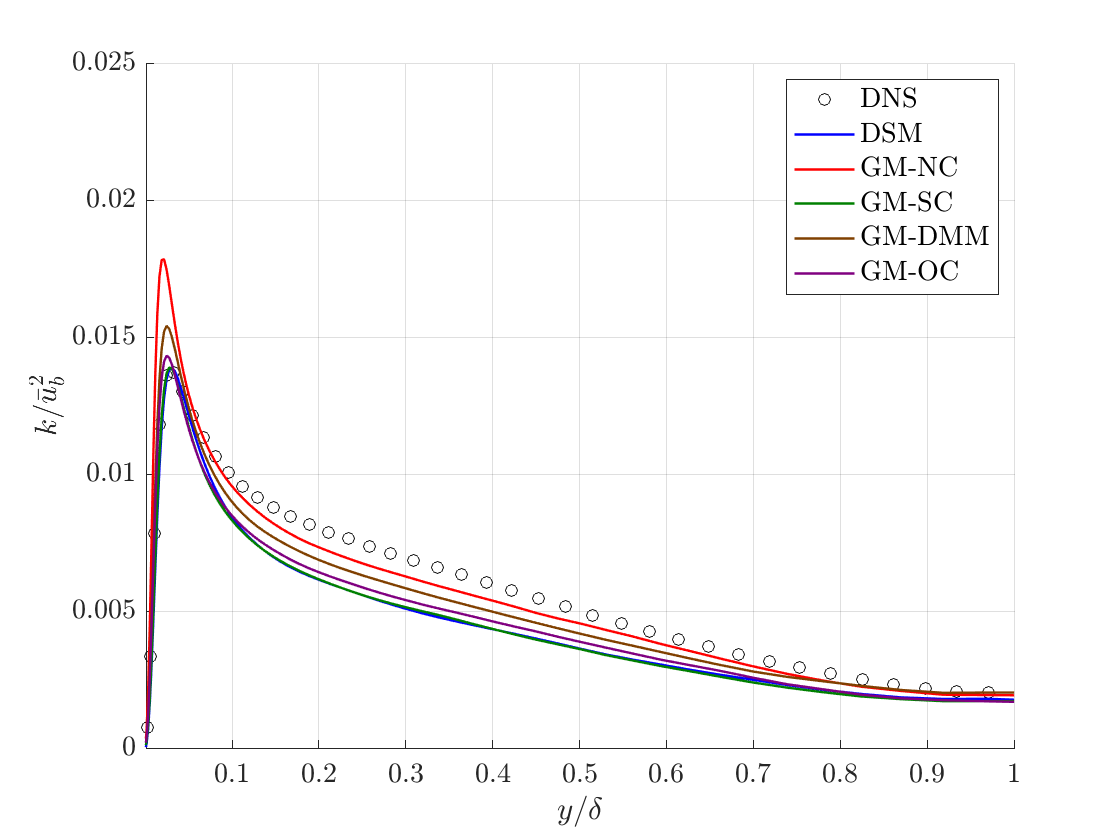}}
    \subfigure[\label{fig:590_a12_medium}]{\includegraphics[width=0.49\textwidth]{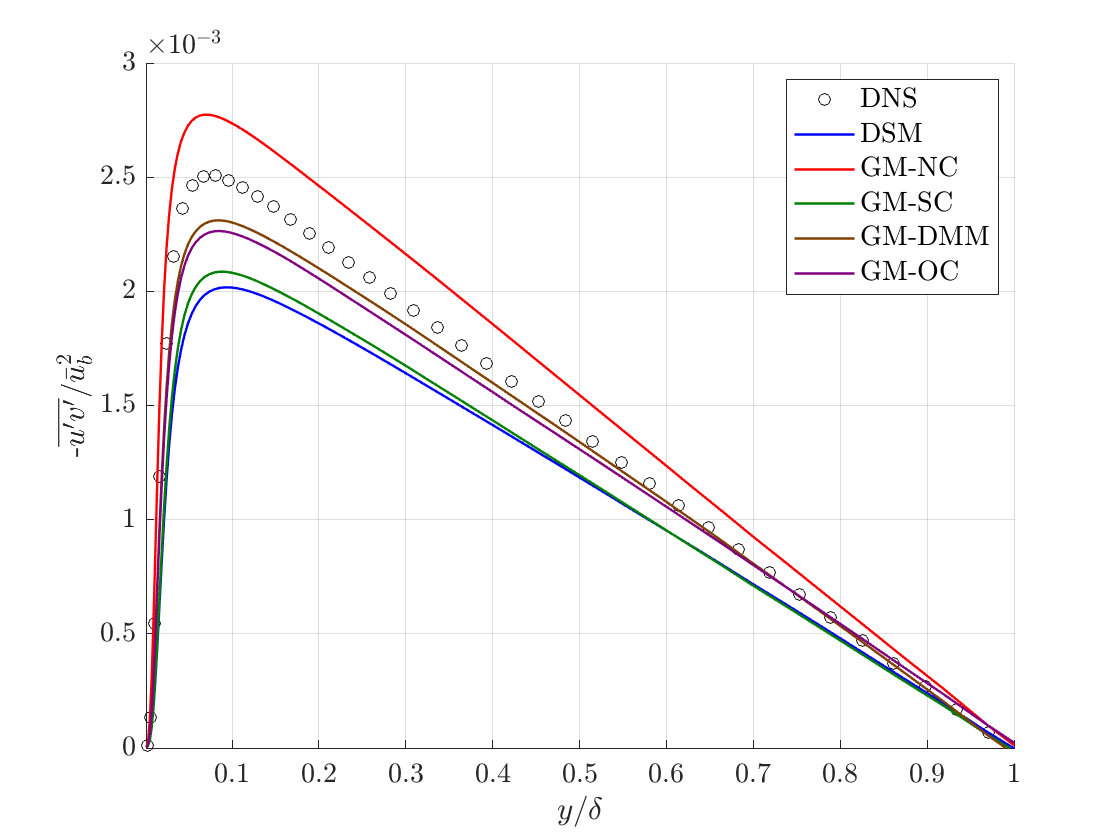}}   
    \caption{ (a) Turbulent kinetic energy and (b) Reynolds shear stress for the medium mesh case for turbulent channel flow at $Re_{\tau} = 590$}
    \label{fig:Channel_590_medium}
\end{figure}

Turbulent kinetic energy and Reynolds shear stress predictions are shown in \figref{Channel_590_coarse} and \figref{Channel_590_medium} for the coarse and medium mesh resolutions respectively. Reynolds stresses are computed by adding the resolved Reynolds stresses and time-averaged subgrid stresses. We observe that the gradient model without clipping significantly overpredicts the peak turbulent kinetic energy for both mesh resolutions. The dynamic Smagorinsky model, the gradient model with standard clipping, the gradient model with optimal clipping and the dynamic mixed model slightly over-predict the peak of turbulent kinetic energy for the coarse mesh resolution, whereas the fine mesh resolution predictions for these models are close to the DNS turbulent kinetic energy. We also observe that the gradient model without clipping also overpredicts the Reynolds shear stress. On the other hand, the dynamic Smagorinsky model and the gradient model without clipping significantly underpredict the Reynolds shear stress. The gradient model with optimal clipping and the dynamic mixed model give much better predictions of Reynolds shear stress compared to the other models, with the dynamic mixed model giving slightly better results than the gradient model with optimal clipping.


The slightly better predictions attained by the dynamic mixed model as compared with the optimally clipped gradient model for turbulent channel flow come at the cost of increased model evaluation cost and implementation complexity. This increased cost is due to the need for explicit filtering and averaging over homogeneous directions at each time step for the dynamic mixed model. The optimally clipped gradient model, on the other hand, does not require these expensive operations. The above results also indicate that optimal clipping leads to a better velocity prediction, peak turbulent kinetic energy prediction, and Reynolds shear stress prediction as compared to standard clipping for turbulent channel flow at $Re_{\tau} = 590$. As \textit{a priori} model dissipation estimates for these two models are the same, we believe that improved structural accuracy is the reason for the enhanced performance of the optimally clipped gradient model as compared to the gradient model with standard clipping. 

\section{Conclusions}
\label{sec:conclusions}

In this article, we proposed a clipping method for structural SGS tensor closures for large eddy simulation. The clipping method results from minimizing the Frobenius norm of the difference between the clipped SGS tensor and the tensor predicted by the structural model subject to the constraint of no local model backscatter. As this method maintains as accurate SGS tensor prediction as possible while preventing local model backscatter, we refer to this method as optimal clipping. Mathematically, we show that optimal clipping corresponds to a particular mixed model formulation. Compared to the dynamic mixed model formulation, the optimal clipping model form is less expensive to evaluate as it does not involve additional filtering and homogeneous direction averaging steps. 

We performed \textit{a priori} and \textit{a posteriori} tests and compared the performance of Clark's gradient SGS model with optimal clipping to that of other models such as the dynamic Smagorinsky model, Clark's gradient model without clipping, Clark's gradient model with standard clipping, and a dynamic mixed model combining Clark's gradient model with a Smagorsinsky term. \textit{A priori} tests indicated that the gradient model with standard clipping reduces the correlation between the predicted SGS tensor and the exact SGS tensor as compared to the gradient model without clipping. This reduction in correlation is aggravated with increased anisotropy in the filtering. Alternatively, the optimally clipped gradient model does not reduce the correlation between the predicted SGS tensor and the exact SGS tensor for both isotropic and anisotropic filtering. 

We performed \textit{a posteriori} tests for three different cases: forced HIT at a high Reynolds number, Taylor-Green vortex flow at $Re = 1600$, and turbulent channel flow at $ Re_{\tau} = 590$. Simulations for forced HIT flow indicated that the use of no explicit SGS model and Clark's gradient model without clipping at high Reynolds numbers results in an energy pileup at large wavenumbers. Both standard and optimal clipping effectively reduced energy pileup. Results from Taylor-Green vortex flow at $Re = 1600$ indicated that the dynamic mixed model leads to a significant underprediction of turbulent kinetic energy and resolved dissipation, while the standard and optimally clipped gradient models both yield accurate results for these quantities. This suggests that clipping may yield more accurate results than the dynamic mixed model for flows exhibiting laminar-turbulent transition. For these first two cases, optimal clipping yielded slightly improved results as compared with standard clipping but the improvements were not significant. However, turbulent channel flow simulations indicated that optimal clipping may yield significantly improved results over standard clipping for wall-resolved LES of wall-bounded flows. For this test case, optimal clipping yielded better mean velocity predictions, peak turbulent kinetic energy predictions and Reynolds shear stress predictions for turbulent channel flow at $Re_{\tau} = 590$ for both coarse and medium mesh resolutions. We hypothesize that even though standard and optimal clipping yield the same \textit{a priori} model dissipation, the improvement in structural accuracy that is attained with optimal clipping is likely responsible for the improvement in \textit{a posteriori} statistics for turbulent channel flow. Therefore, for flows where clipping is necessary for the stability of \textit{a posteriori} simulations,  optimal clipping should be selected over standard clipping as it preserves the structural accuracy of the predicted model stresses of the underlying structural SGS model.

Although all results shown here were attained with Clark's gradient model as the base structural SGS model, optimal clipping can be applied to other structural SGS models such as Bardina's scale-similarity model, approximate deconvolution models, and data-driven models without modification. In fact, optimal clipping was also tested with the Bardina's scale-similarity \cite{Bardina1980} and the data-driven model presented in \cite{Prakash2021}, and similar trends to those reported here were also observed for these models. The significant improvement in turbulent channel flow predictions observed here is a motivation for testing optimal clipping for other complicated wall-bounded flows such as a developing turbulent boundary layer, a boundary layer subject to favorable and/or adverse pressure gradients, or even a separating boundary layer.

\section*{Funding Sources}
This research was made possible using the funds from Computational and Data-Enabled Science and Engineering (CDS\&E) program of the National Science Foundation (NSF) CBET-1710670, as well as the Transformational Tools and Technologies Project of the National Aeronautics and Space Administration (NASA) 80NSSC18M0147.

\section*{Acknowledgments}
The authors would like to thank the Argonne Leadership Computing Facility (ALCF) for the resources on which the simulations and post-processing were performed.

\bibliography{main.bib}

\begin{thebibliography}{37}
\newcommand{\enquote}[1]{``#1''}
\providecommand{\natexlab}[1]{#1}
\providecommand{\url}[1]{\texttt{#1}}
\providecommand{\urlprefix}{URL }
\expandafter\ifx\csname urlstyle\endcsname\relax
  \providecommand{\doi}[1]{\discretionary{}{}{}https://doi.org/#1}\else
  \providecommand{\doi}[1]{\discretionary{}{}{}\urlstyle{rm}\url{https://doi.org/#1}}\fi

\bibitem[{Chapman(1979)}]{Chapman1979}
Chapman, D.~R., \enquote{Computational aerodynamics development and outlook,}
  \emph{AIAA Journal}, Vol.~17, No.~12, 1979, pp. 1293--1313.
\newblock \doi{10.2514/3.61311}.

\bibitem[{Choi and Moin(2012)}]{choi2012}
Choi, H., and Moin, P., \enquote{Grid-point requirements for large eddy
  simulation: Chapman’s estimates revisited,} \emph{Physics of Fluids},
  Vol.~24, No.~1, 2012, p. 011702.
\newblock \doi{10.1063/1.3676783}.

\bibitem[{Smagorinsky(1963)}]{Smagorinsky1963}
Smagorinsky, J., \enquote{General circulation experiments with the primitive
  equations,} \emph{Monthly Weather Review}, Vol.~91, No.~3, 1963, pp. 99--164.
\newblock \doi{10.1175/1520-0493(1963)091<0099:GCEWTP>2.3.CO;2}.

\bibitem[{Germano et~al.(1991)Germano, Piomelli, Moin, and Cabot}]{Germano1991}
Germano, M., Piomelli, U., Moin, P., and Cabot, W.~H., \enquote{{A dynamic
  subgrid-scale eddy viscosity model},} \emph{Physics of Fluids A}, Vol.~3,
  No.~7, 1991, pp. 1760--1765.
\newblock \doi{10.1063/1.857955}.

\bibitem[{Clark et~al.(1979)Clark, Ferziger, and Reynolds}]{Clark1979}
Clark, R.~A., Ferziger, J.~H., and Reynolds, W.~C., \enquote{Evaluation of
  subgrid-scale models using an accurately simulated turbulent flow,}
  \emph{Journal of Fluid Mechanics}, Vol.~91, No.~1, 1979, p. 1–16.
\newblock \doi{10.1017/S002211207900001X}.

\bibitem[{Bardina et~al.(1980)Bardina, Ferziger, and Reynolds}]{Bardina1980}
Bardina, J., Ferziger, J.~H., and Reynolds, W.~C., \enquote{{Improved
  subgrid-scale models for large-eddy simulation.}} \emph{AIAA Paper}, 1980.
\newblock \doi{10.2514/6.1980-1357}.

\bibitem[{Stolz et~al.(2001)Stolz, Adams, and Kleiser}]{Stolz2001}
Stolz, S., Adams, N.~A., and Kleiser, L., \enquote{An approximate deconvolution
  model for large-eddy simulation with application to incompressible
  wall-bounded flows,} \emph{Physics of Fluids}, Vol.~13, No.~4, 2001, pp.
  997--1015.
\newblock \doi{10.1063/1.1350896}.

\bibitem[{Borue and Orszag(1998)}]{Borue1998}
Borue, V., and Orszag, S.~A., \enquote{{Local energy flux and subgrid-scale
  statistics in three-dimensional turbulence},} \emph{Journal of Fluid
  Mechanics}, Vol. 366, 1998, pp. 1--31.
\newblock \doi{10.1017/S0022112097008306}.

\bibitem[{Liu et~al.(1994)Liu, Meneveau, and Katz}]{Liu1994}
Liu, S., Meneveau, C., and Katz, J., \enquote{On the properties of similarity
  subgrid-scale models as deduced from measurements in a turbulent jet,}
  \emph{Journal of Fluid Mechanics}, Vol. 275, 1994, p. 83–119.
\newblock \doi{10.1017/S0022112094002296}.

\bibitem[{Vreman et~al.(1997)Vreman, Geurts, and Kuerten}]{Vreman1997}
Vreman, B., Geurts, B., and Kuerten, H., \enquote{Large-eddy simulation of the
  turbulent mixing layer,} \emph{Journal of Fluid Mechanics}, Vol. 339, 1997,
  p. 357–390.
\newblock \doi{10.1017/S0022112097005429}.

\bibitem[{Vreman et~al.(1996)Vreman, Geurts, and Kuerten}]{Vreman1996}
Vreman, B., Geurts, B., and Kuerten, H., \enquote{{Large-eddy simulation of the
  temporal mixing layer using the Clark model},} \emph{Theoretical and
  Computational Fluid Dynamics}, Vol.~8, No.~4, 1996, pp. 309--324.
\newblock \doi{10.1007/BF00639698}.

\bibitem[{Zang et~al.(1993)Zang, Street, and Koseff}]{Zang1993}
Zang, Y., Street, R.~L., and Koseff, J.~R., \enquote{A dynamic mixed
  subgrid‐scale model and its application to turbulent recirculating flows,}
  \emph{Physics of Fluids A: Fluid Dynamics}, Vol.~5, No.~12, 1993, pp.
  3186--3196.
\newblock \doi{10.1063/1.858675}.

\bibitem[{Vollant et~al.(2016)Vollant, Balarac, and Corre}]{Vollant2016}
Vollant, A., Balarac, G., and Corre, C., \enquote{A dynamic regularized
  gradient model of the subgrid-scale stress tensor for large-eddy simulation,}
  \emph{Physics of Fluids}, Vol.~28, No.~2, 2016, p. 025114.
\newblock \doi{10.1063/1.4941781}.

\bibitem[{Baggett et~al.(1997)Baggett, Jim{\'{e}}nez, and
  Kravchenko}]{Baggett1997}
Baggett, J.~S., Jim{\'{e}}nez, J., and Kravchenko, A.~G., \enquote{{Resolution
  requirements in large-eddy simulations of shear flows},} \emph{Center for
  Turbulence Research Annual Research Briefs}, 1997, pp. 51--66.

\bibitem[{Deardorff(1970)}]{Deardorff1970}
Deardorff, J.~W., \enquote{A numerical study of three-dimensional turbulent
  channel flow at large Reynolds numbers,} \emph{Journal of Fluid Mechanics},
  Vol.~41, No.~2, 1970, p. 453–480.
\newblock \doi{10.1017/S0022112070000691}.

\bibitem[{Lilly(1992)}]{Lilly1992}
Lilly, D.~K., \enquote{{A proposed modification of the Germano subgrid-scale
  closure method},} \emph{Physics of Fluids A}, Vol.~4, No.~3, 1992, pp.
  633--635.
\newblock \doi{10.1063/1.858280}.

\bibitem[{Canuto and Cheng(1997)}]{Canuto1997}
Canuto, V.~M., and Cheng, Y., \enquote{Determination of the Smagorinsky–Lilly
  constant $C_S$,} \emph{Physics of Fluids}, Vol.~9, No.~5, 1997, pp.
  1368--1378.
\newblock \doi{10.1063/1.869251}.

\bibitem[{Germano(1992)}]{Germano1992}
Germano, M., \enquote{Turbulence: The filtering approach,} \emph{Journal of
  Fluid Mechanics}, Vol. 238, 1992, p. 325–336.
\newblock \doi{10.1017/S0022112092001733}.

\bibitem[{Rozema et~al.(2015)Rozema, Bae, Moin, and Verstappen}]{Rozema2015}
Rozema, W., Bae, H.~J., Moin, P., and Verstappen, R.,
  \enquote{Minimum-dissipation models for large-eddy simulation,} \emph{Physics
  of Fluids}, Vol.~27, No.~8, 2015, p. 085107.
\newblock \doi{10.1063/1.4928700}.

\bibitem[{Peters et~al.(2022)Peters, Balin, Jansen, Doostan, and
  Evans}]{Peters2020}
Peters, E.~L., Balin, R., Jansen, K.~E., Doostan, A., and Evans, J.~A.,
  \enquote{S-frame discrepancy correction models for data-informed Reynolds
  stress closure,} \emph{Journal of Computational Physics}, Vol. 448, 2022, p.
  110717.
\newblock \doi{10.1016/j.jcp.2021.110717}.

\bibitem[{Silvis et~al.(2017)Silvis, Remmerswaal, and Verstappen}]{Silvis2017}
Silvis, M.~H., Remmerswaal, R.~A., and Verstappen, R., \enquote{Physical
  consistency of subgrid-scale models for large-eddy simulation of
  incompressible turbulent flows,} \emph{Physics of Fluids}, Vol.~29, No.~1,
  2017, p. 015105.
\newblock \doi{10.1063/1.4974093}.

\bibitem[{Prakash et~al.(2021)Prakash, Jansen, and Evans}]{Prakash2021}
Prakash, A., Jansen, K.~E., and Evans, J.~A., \enquote{Invariant data-driven
  subgrid stress modeling in the strain-rate eigenframe for large eddy
  simulation,} \emph{arXiv:2106.13410}, 2021.
\newblock \urlprefix\url{https://arxiv.org/abs/2106.13410}.

\bibitem[{Trias et~al.(2017)Trias, Gorobets, Silvis, Verstappen, and
  Oliva}]{Trias2017}
Trias, F.~X., Gorobets, A., Silvis, M.~H., Verstappen, R. W. C.~P., and Oliva,
  A., \enquote{A new subgrid characteristic length for turbulence simulations
  on anisotropic grids,} \emph{Physics of Fluids}, Vol.~29, No.~11, 2017, p.
  115109.
\newblock \doi{10.1063/1.5012546}.

\bibitem[{Li et~al.(2008)Li, Perlman, Wan, Yang, Meneveau, Burns, Chen, Szalay,
  and Eyink}]{Li2008}
Li, Y., Perlman, E., Wan, M., Yang, Y., Meneveau, C., Burns, R., Chen, S.,
  Szalay, A., and Eyink, G., \enquote{A public turbulence database cluster and
  applications to study Lagrangian evolution of velocity increments in
  turbulence,} \emph{Journal of Turbulence}, Vol.~9, 2008, p. N31.
\newblock \doi{10.1080/14685240802376389}.

\bibitem[{PHA(2021)}]{PHASTA}
\enquote{{PHASTA},} \url{https://github.com/PHASTA/phasta}, 2021.
\newblock Accessed: 2021-06-21.

\bibitem[{Whiting and Jansen(2001)}]{Whiting2001}
Whiting, C.~H., and Jansen, K.~E., \enquote{A stabilized finite element method
  for the incompressible Navier–Stokes equations using a hierarchical basis,}
  \emph{International Journal for Numerical Methods in Fluids}, Vol.~35, No.~1,
  2001, pp. 93--116.
\newblock \doi{10.1002/1097-0363(20010115)35:1<93::AID-FLD85>3.0.CO;2-G}.

\bibitem[{Jansen et~al.(2000)Jansen, Whiting, and Hulbert}]{Jansen2000}
Jansen, K.~E., Whiting, C.~H., and Hulbert, G.~M., \enquote{A
  generalized-$\alpha$ method for integrating the filtered {N}avier–{S}tokes
  equations with a stabilized finite element method,} \emph{Computer Methods in
  Applied Mechanics and Engineering}, Vol. 190, No.~3, 2000, pp. 305 -- 319.
\newblock \doi{https://doi.org/10.1016/S0045-7825(00)00203-6},
  \urlprefix\url{http://www.sciencedirect.com/science/article/pii/S0045782500002036}.

\bibitem[{Tejada-Mart{\'{i}}nez and Jansen(2003)}]{Tejada2003}
Tejada-Mart{\'{i}}nez, A.~E., and Jansen, K.~E., \enquote{Spatial test filters
  for dynamic model large-eddy simulation with finite elements,}
  \emph{Communications in Numerical Methods in Engineering}, Vol.~19, No.~3,
  2003, pp. 205--213.
\newblock \doi{https://doi.org/10.1002/cnm.509}.

\bibitem[{Tejada-Mart{\'{i}}nez and Jansen(2005)}]{Tejada2005}
Tejada-Mart{\'{i}}nez, A.~E., and Jansen, K.~E., \enquote{On the interaction
  between dynamic model dissipation and numerical dissipation due to streamline
  upwind/Petrov–Galerkin stabilization,} \emph{Computer Methods in Applied
  Mechanics and Engineering}, Vol. 194, No.~9, 2005, pp. 1225--1248.
\newblock \doi{https://doi.org/10.1016/j.cma.2004.06.037}.

\bibitem[{Trofimova et~al.(2009)Trofimova, Tejada-Mart{\'{i}}nez, Jansen, and
  Lahey}]{Trofimova2009}
Trofimova, A.~V., Tejada-Mart{\'{i}}nez, A.~E., Jansen, K.~E., and Lahey,
  R.~T., \enquote{Direct numerical simulation of turbulent channel flows using
  a stabilized finite element method,} \emph{Computers \& Fluids}, Vol.~38,
  No.~4, 2009, pp. 924--938.
\newblock \doi{10.1016/j.compfluid.2008.10.003}.

\bibitem[{Balin and Jansen(2021)}]{Balin2021}
Balin, R., and Jansen, K., \enquote{Direct numerical simulation of a turbulent
  boundary layer over a bump with strong pressure gradients,} \emph{Journal of
  Fluid Mechanics}, Vol. 918, 2021, p. A14.
\newblock \doi{10.1017/jfm.2021.312}.

\bibitem[{Bazilevs et~al.(2007)Bazilevs, Calo, Cottrell, Hughes, Reali, and
  Scovazzi}]{Bazilevs2007}
Bazilevs, Y., Calo, V., Cottrell, J., Hughes, T., Reali, A., and Scovazzi, G.,
  \enquote{Variational multiscale residual-based turbulence modeling for large
  eddy simulation of incompressible flows,} \emph{Computer Methods in Applied
  Mechanics and Engineering}, Vol. 197, No.~1, 2007, pp. 173 -- 201.
\newblock \doi{10.1016/j.cma.2007.07.016}.

\bibitem[{Pope(2000)}]{Pope2000}
Pope, S.~B., \emph{Turbulent Flows}, Cambridge University Press, 2000.
\newblock \doi{10.1017/CBO9780511840531}.

\bibitem[{Shoraka(2017)}]{Shoraka2017}
Shoraka, Y., \enquote{Large-eddy simulation of turbulence and turbulent mixing
  in compressible turbulent flows without and with shear,} Ph.D. thesis,
  University of New South Wales, Australia, 2017.

\bibitem[{Brachet et~al.(1983)Brachet, Meiron, Orszag, Nickel, Morf, and
  Frisch}]{Brachet1983}
Brachet, M.~E., Meiron, D.~I., Orszag, S.~A., Nickel, B.~G., Morf, R.~H., and
  Frisch, U., \enquote{Small-scale structure of the Taylor–Green vortex,}
  \emph{Journal of Fluid Mechanics}, Vol. 130, 1983, p. 411–452.
\newblock \doi{10.1017/S0022112083001159}.

\bibitem[{Germano(1986)}]{Germano1986}
Germano, M., \enquote{Differential filters for the large eddy numerical
  simulation of turbulent flows,} \emph{The Physics of Fluids}, Vol.~29, No.~6,
  1986, pp. 1755--1757.
\newblock \doi{10.1063/1.865649}.

\bibitem[{Moser et~al.(1999)Moser, Kim, and Mansour}]{Moser1999}
Moser, R.~D., Kim, J., and Mansour, N.~N., \enquote{{Direct numerical
  simulation of turbulent channel flow up to $Re_{\tau} = 590$},} \emph{Physics
  of Fluids}, Vol.~11, No.~4, 1999, pp. 943--945.
\newblock \doi{10.1063/1.869966}.

\end{thebibliography}

\end{document}